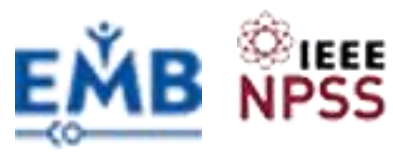
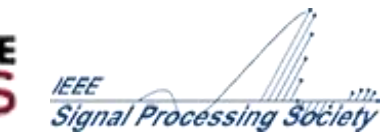

# Phase-Locked Time-Stretch Optical Coherence Tomography for Contrast-Enhanced Retinal Microangiography

Gyeong Hun Kim, Seongjin Bak, Hyung-Hoi Kim, Jun Geun Shin, Tae Joong Eom, Chang-Seok Kim and Hwidon Lee

***Abstract*—Optical coherence tomography angiography has transformed retinal vascular imaging by providing non-invasive, high-resolution visualization. However, achieving an optimal balance between field of view, resolution, and three-dimensional microvasculature contrast, particularly in deeper retinal layers, remains challenging. A phase-locked time-stretch optical coherence tomography microangiography system is developed to address these limitations with a 5-MHz A-line rate and sub-nm phase sensitivity. Utilizing a dual chirped fiber Bragg grating architecture, the swept-source laser achieves an extended coherence length of ~10 mm and a 102-nm bandwidth. A time-stretch analog-to-digital converter overcomes the limitations of conventional multi-MHz optical coherence tomography systems, ensuring a 2-mm imaging depth in the air with high spatial resolution. The proposed system enables high-contrast, depth-encoded mapping of key retinal structures, including the superficial and deep capillary plexuses and the choriocapillaris. Compared to a state-of-the-art system, the proposed approach demonstrates enhanced resolution, improved contrast, and faster imaging speeds, enhancing its potential for diagnosing and monitoring retinal and systemic diseases like age-related macular degeneration, diabetic retinopathy, and Alzheimer's disease.**

***Index Terms*— Optical coherence tomography angiography (OCTA), Retinal microvasculature, Stretched-pulse active mode-locking (SPML), Swept source**

## I. Introduction

HIGH-resolution imaging of retinal blood vessels is crucial for the early detection and monitoring of retinal and systemic diseases that affect vascular health. Optical coherence tomography angiography (OCTA) has emerged as a revolutionary imaging modality to visualize retinal vasculature. Unlike conventional techniques that require contrast dyes, OCTA leverages the intrinsic motion contrast of blood cells, offering a non-invasive imaging solution [1]. Compared with traditional OCTA methods, the approach based on complex variance (CV) demonstrates superior noise suppression and enhances the contrast between flowing blood and stationary tissues [2-4]. CV is a phase-sensitive optical coherence tomography (OCT) technique that incorporates both intensity and phase information from OCT datasets, thereby enabling the detection of subtle vascular changes with high clarity. Moreover, CV-based OCTA facilitates three-dimensional (3D) segmentation and mapping of the microvasculature across individual retinal layers, offering a comprehensive assessment of vascular health.

Although CV-based OCTA is a powerful technique, it requires repeated cross-sectional imaging of identical locations at different times while maintaining high phase coherency among the complex-valued data. However, this is challenging

This paragraph of the first footnote will contain the date on which you submitted your paper for review.

This work was supported by the National Research Foundation of Korea (NRF) grant funded by the Korea Government (MSIT) (No. NRF-2021R1A5A1032937) and a grant of the Korea Health Technology R&D Project through the Korea Health Industry Development Institute (KHIDI), funded by the Ministry of Health & Welfare, Republic of Korea (No. HR20C0026).

This study involving human subjects was approved by the Institutional Review Board of Pusan National University (IRB Protocol No. PNU IRB/2023_07_HR) and conducted in accordance with ethical guidelines and the Declaration of Helsinki.

Gyeong Hun Kim was with the Engineering Research Center for Color-Modulated Extra-Sensory Perception Technology, Pusan National University, Busan, 46241, Republic of Korea. He is now with the Wellman Center for Photomedicine, Massachusetts General Hospital, Harvard Medical School, Boston, MA 02114 USA (e-mail: gkim20@mgh.harvard.edu).

Seongjin Bak was with the Engineering Research Center for Color-Modulated Extra-Sensory Perception Technology, Pusan National University, Busan, 46241, Republic of Korea, and the Department of Cogno-Mechatronics Engineering, Pusan National University, Busan, 46241, Republic of Korea (e-mail: sj.bak@pusan.ac.kr).

Hyung-Hoi Kim was with the Department of Laboratory Medicine and Biomedical Research Institute, Pusan National University Hospital, Busan, 49241, Republic of Korea (email: hhkim@pusan.ac.kr).

Jun Geun Shin was with the Optical Precision Measurement Research Center, Korea Photonics Technology Institute, Gwangju, 61007, Republic of Korea (e-mail: jgshin@gist.ac.kr).

Tae Joong Eom is with the Engineering Research Center for Color-Modulated Extra-Sensory Perception Technology, Pusan National University, Busan, 46241, Republic of Korea, and the Department of Cogno-Mechatronics Engineering, Pusan National University, Busan, 46241, Republic of Korea (e-mail: eomtj@pusan.ac.kr).

Chang-Seok Kim is with the Engineering Research Center for Color-Modulated Extra-Sensory Perception Technology, Pusan National University, Busan, 46241, Republic of Korea, and the Department of Cogno-Mechatronics Engineering, Pusan National University, Busan, 46241, Republic of Korea (e-mail: ckim@pusan.ac.kr).

Hwidon Lee is with the Engineering Research Center for Color-Modulated Extra-Sensory Perception Technology, Pusan National University, Busan, 46241, Republic of Korea, and the Department of Cogno-Mechatronics Engineering, Pusan National University, Busan, 46241, Republic of Korea (e-mail: hwidonlee@pusan.ac.kr).

because of the slow imaging speed of spectral-domain OCT, which has become more pronounced in OCTA applications [4]. These slow speeds increase the risk of image distortions caused by involuntary patient movements such as blinking or saccades [5]. To address these issues, swept-source (SS) OCT systems offering rapid tuning speeds, wide bandwidths, and extended coherence lengths have been developed [6]. Cutting-edge SS technologies, such as Fourier domain mode-locking (FDML) and microelectromechanical system (MEMS)-tunable vertical-cavity surface-emitting lasers (VCSELs), enable high-speed 3D OCT imaging [7-15]. FDML lasers can achieve fundamental sweep rates of up to 440 kHz by synchronizing the fiber laser cavity with the driving frequency of a wavelength-tunable filter [6]. Conversely, the thermal stresses and piezoelectric crystal responses can constrain these sweep rates. The sweep rates can be increased to 3.5 MHz using a buffering technique; however, this approach may introduce spectral variances owing uneven optical separation and path length differences [10, 11]. MEMS-tunable VCSELs enable high-speed, long-range OCT imaging with narrow instantaneous linewidth ideal for meter-range imaging [9, 12, 13]. Nevertheless, they are hindered by vibration modes and thermally induced noise, particularly at high sweep rates [14, 16].

Although SS-OCT systems achieve rapid tuning speeds and wide bandwidths, maintaining high phase stability is challenging because of phase fluctuations and A-line trigger jitters, particularly at high sweep rates [9, 11, 14, 15, 17, 18]. Recently, SSs based on the stretched-pulse active mode-locking (SPML) configuration have shown promise for ultra-high-speed (multi-MHz) and stable operation with wavenumber linearity [19-22]. The repeated stretching, amplification, and compression of the optical pulse in the SPML laser cavity resulted in a long and continuous chirped pulse with a broad spectral bandwidth. All-fiber laser cavities and precise pulse modulation further enhance the mode-locking stability and synchronization with other devices.

Despite these advancements, OCT systems using stretched-pulse active mode-locked swept source (SPML-SS) still face significant challenges, including limited coherence lengths, narrow bandwidths, and shallow imaging depths. These limitations are primarily due to the intrinsic dispersion in optical fibers and the constraints of traditional optical data sampling techniques. Previous studies on SPML laser techniques utilized negative-dispersion-slope optical fibers, such as dispersion-compensating fibers (DCF) and photonic-crystal fibers (PCF), for dispersion compensation [19, 20]. However, the dispersion characteristics of these fibers restrict optimal compensation for high-order dispersions across a wide spectral range, leading to bandwidth and coherence length limitations that restrict the full potential of the SPML technique. Although the 1.3-μm band exhibits comparatively enhanced dispersion characteristics, the residual dispersion in the laser cavity can still limit coherence length and bandwidth, adversely affecting SS-OCT imaging performance [21, 22]. Crucially, the 1.3-μm wavelength is more prone to water attenuation than the 1070-nm range, resulting in lower penetration and reduced imaging quality in *in vivo* human retinal applications [23]. Moreover, the increased sweep rates of the SPML laser, which can reach several MHz, also present another limitation—as imaging speeds increase, the achievable imaging depth decreases and is limited to a ~1-mm range in the air [20-22, 24]. This tradeoff between the imaging speed and the depth is intrinsically linked to the bandwidth limitations of analog-to-digital converters (ADCs) and photodetectors [25]. Traditional optical-data sampling methods that rely on both ADCs and photodetectors cannot provide the required rapid real-time optical measurements.

A dual chirped fiber Bragg grating (CFBG) architecture optimized for 1.0-μm SPML-SS is proposed to overcome these challenges. This system achieved a coherence length of ~10 mm and a bandwidth of 102 nm. Implementing a phase-locked time-stretch technique in SPML-SS achieves sub-nanometer phase stability, 5 MHz imaging speed, and 2 mm imaging depth with high spatial resolution, effectively resolving the long-standing trade-offs among imaging speed, depth, and resolution. Additionally, the fast and stable imaging capabilities enhance the performance of the CV algorithm, improving noise suppression and enabling detailed, depth-resolved 3D mapping of the retinal microvasculature. These advancements have enhanced retinal OCT microangiography (OCTMA), facilitating the precise visualization of vascular networks and advancing the understanding of retinal and systemic vascular diseases.

## II. Methods

### A. Phase-Locked Time-Stretch Optical Coherence Tomography

In an OCT system, the amplitude of each modulation frequency correlates with the scattering profile along the imaging depth. Consequently, the performance parameters of the ADC, such as the sampling rate and jitters of the clock and trigger, directly influence the quality of the OCT images, affecting the imaging depth, noise, and phase sensitivity.

As shown in Fig. 1a, the SPML laser pulse, which initially experienced a linear wavenumber stretch within the laser cavity, underwent a secondary optical time stretch within an external time stretcher. Intensity modulation occurs along the wavenumber when the stretched pulse travels through the optical interferometer. Synchronizing the SPML-SS and ADC with a rubidium atomic clock timebase yields high temporal precision and long-term frequency stability below 1 ppb/year. Under phase-locked conditions, this configuration enables accurate spectral sampling and direct digitization along the wavenumber axis, as shown in Fig. 1b. The time-stretch mechanism significantly enhanced the spectral data sampling rate, effectively doubling it compared to a regular ADC. This improvement enabled an imaging depth of 2 mm in air (Fig. 1c).

A compilation of 1000 interference fringes captured in the phase-locked condition using a common-path interferometer (CPI) is shown in Fig. 1d. The CPI was equipped with a 150-μm commercial cover glass, resulting in an optical path difference of 465 μm. The magnified inset highlights the exceptional spectral sampling stability achieved through phase-locking. The phase stability calculated from the phase angles at the frequency-domain peaks of the interference fringes, is shown in Fig. 1e. Under the phase-locked condition, the phase stability was 4.8 mrad with a signal-to-noise ratio (SNR) of 44 dB, corresponding to a displacement sensitivity of 0.31 nm, representing a 6.7% difference compared to the theoretical phase stability of 4.5 mrad expected under shot-noise limitation at the same SNR [18]. In contrast, significantly higher variability was observed in the unlocked condition.

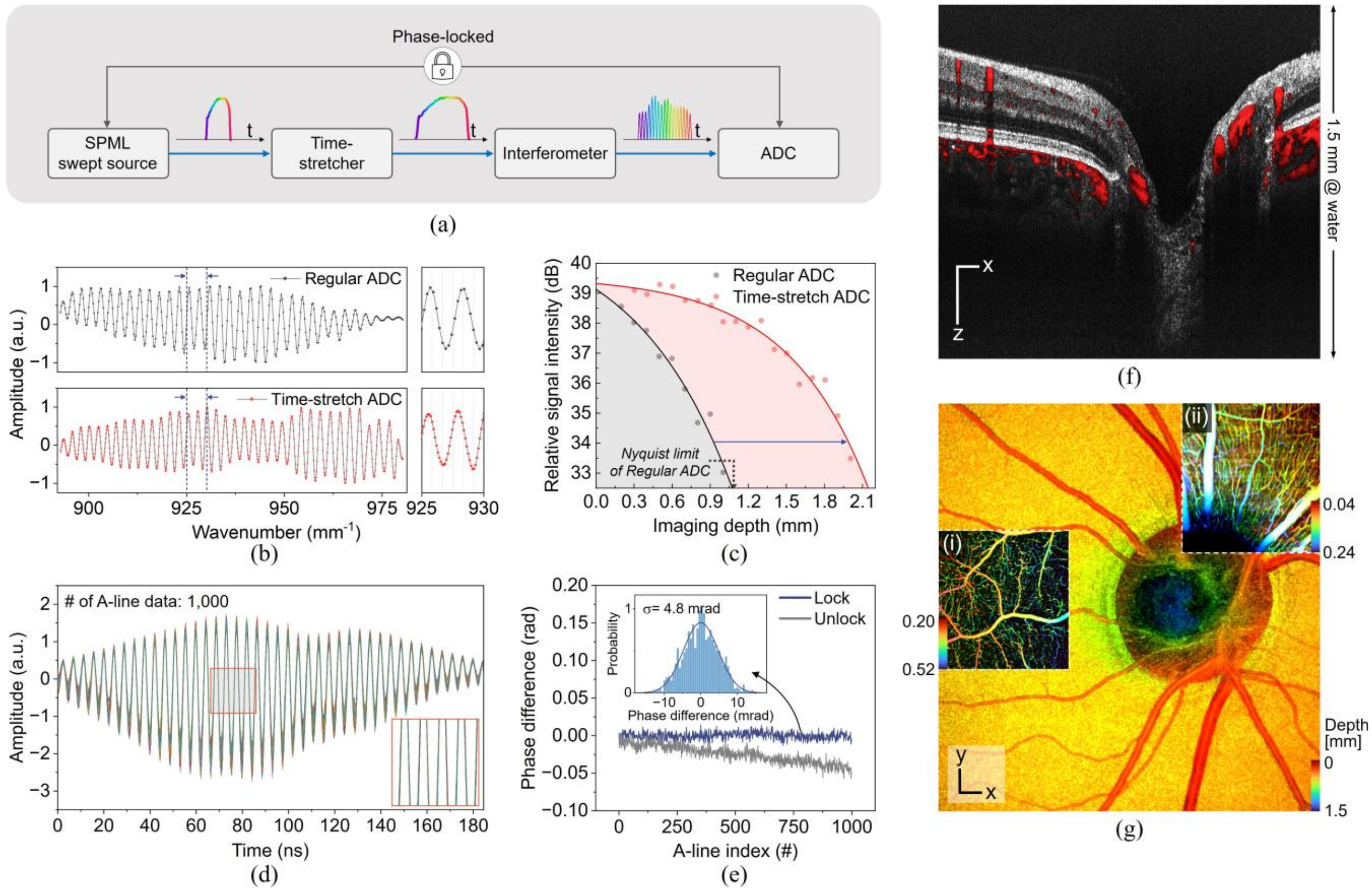

Fig. 1. Concept of phase-locked time-stretch OCT. (a) Diagram illustrating the phase-locked time-stretch method, employing a stretched-pulse active mode-locked swept source (SPML-SS) synchronized with an analog-to-digital converter (ADC) to achieve high-precision spectral sampling. The rainbow colors in the time-domain spectra represent the wavelength components, stretched linearly with respect to the wavenumber by the specially engineered dispersion profile of time stretcher. (b) Interference signals recorded using two different ADCs, showcasing doubled spectral data sampling via the time-stretch mechanism in comparison to regular ADC. (c) Signal intensity roll-off based on the imaging depth, demonstrating enhanced imaging depth (~2 mm in air) due to the time-stretch mechanism. (d) Overlapped display of 1000 interference fringes under phase-locked conditions, with a magnified inset highlighting the extremely stable spectral sampling achieved via rubidium atomic clock synchronization. (e) Phase stability derived from (d), with 4.8-mrad stability (0.31-nm displacement sensitivity) under the phase-locked condition. In contrast, the unlock condition exhibits significantly higher variability. (f) Overlay of retinal OCT cross-sectional structural image (grayscale) and CV-OCTMA image (red scale) in the optic disc area, achieving an imaging depth up to 1.5 mm in an aqueous environment ($n_w$ = 1.33). (g) Depth-encoded maximum intensity projection (MIP) of the 3D structural OCT image overlaid with CV-OCTMA images. The primary panel shows the structural OCT MIP, while insets (i) and (ii) display en-face MIPs of 3D CV-OCTMA images, highlighting the superficial vascular complex and radial peripapillary capillary plexus, respectively. The CV-OCTMA images provide depth-resolved, high-contrast visualization of microvascular networks across the retina. Scale bar = 250 μm.

Fig. 1f shows an overlay of the cross-sectional retinal OCT structural image (grayscale) and the CV-OCTMA image (red scale) of the optic disc region, achieving an imaging depth of 1.5 mm in an aqueous environment with a refractive index ($n_w$) of ~1.33. Fig. 1g shows the depth-encoded maximum intensity projection (MIP) of the 3D structural OCT image, partially overlaid with CV-OCTMA images, highlighting key retinal microvasculature. The primary panel shows the structural OCT MIP, while insets (i) and (ii) display en-face MIPs of 3D CV-OCTMA images. Inset (i) highlights the superficial vascular complex (SVC), derived from data within a depth range of 0.20–0.52 mm, while inset (ii) focuses on the radial peripapillary capillary plexus, based on the data from a depth range of 0.04–0.24 mm. These CV-OCTMA images provide depth-resolved and high-contrast visualization of the retinal microvascular networks.

### B. Dual CFBG Stretched-Pulse Active Mode-Locked Swept Source

The proposed 1.0-μm SPML-SS includes an SPML seed laser based on a dual CFBG architecture alongside an external stage for time-stretching and amplification (Fig. 2a). The seed laser uses a theta (θ)-shaped cavity, generating wideband and ultra-fast stretched pulse trains through the application of a time-stretch and active mode-locking technique. This process encompasses a cycle of stretching, amplification, and compression during each round trip, which is repeated successively. Stable mode locking was ensured by actively modulating the laser gain between the compression and restretching stages using electrical pulse signals with a modulation frequency that precisely matched the fundamental resonance frequency of the laser cavity. A Mach–Zehnder-type electro-optic intensity modulator (EOM, NIR-MX-LN-20, iXblue, France) was used to modulate the gain in the cavity. EOM characterized by a bandwidth of ~20 GHz and an extinction ratio of ~30 dB, can generate a short-pulse optical profile with a width of up to 50 ps. Three optical fiber circulators (CIR, FOC-12N-111-6/125-PPP-1064-50, OZ Optics Ltd., Canada) were employed within the laser cavity to ensure unidirectional optical circulation and serve as inline polarizers for the fast axis blocking. A quantum-dot semiconductor optical amplifier (SOA, SOA-1060-90-PM-30dB, Innolume GmbH, Germany) was implemented as the gain medium, which delivered a wide bandwidth and high small-signal gain (typically 30 dB) of ~100 nm centered at 1.07 μm. A 97-ns electrical pulse was used to synchronously switch the SOA at the modulation frequency, preventing undesired lasing within the laser cavity.

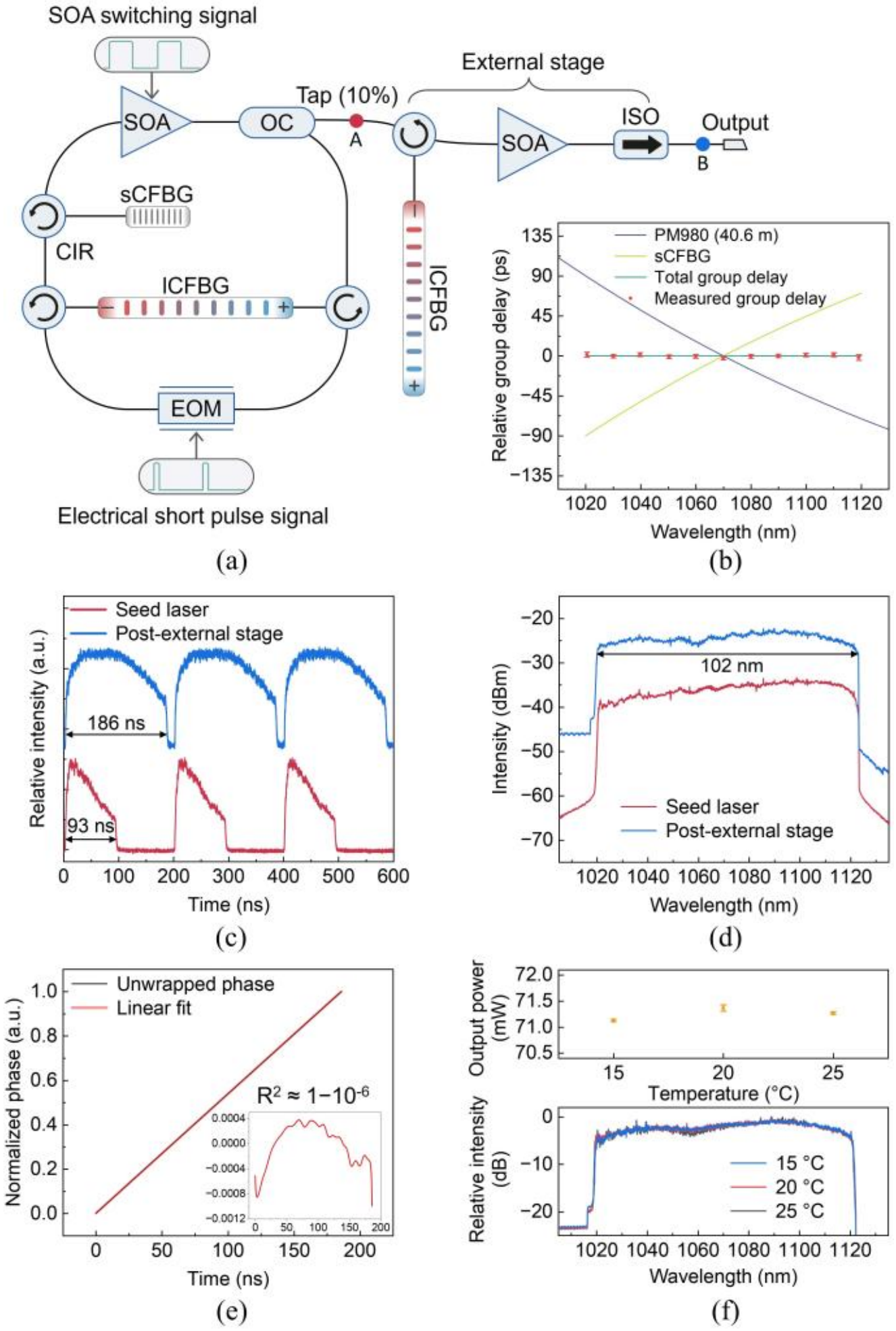

(a) (b) (c) (d) (e) (f)

Fig. 2. Design and performance of the 1.0-μm SPML-SS. (a) Schematic of the dual chirped fiber Bragg grating (CFBG) architecture-based seed laser and additional external stage for time-stretching and amplification. The red circle (Point A) shows the seed output position, and the blue circle (Point B) indicates the post-external stage position. (b) Dispersion curve of the laser cavity measured over a 100-nm bandwidth centered at 1070 nm. The measured group delay (3.8 ps) aligns with the designed dispersion profile, confirming precise dispersion control across the operating spectral range. (c) Time traces of the seed laser and post-external stage outputs of SPML-SS. The stable time-domain signal demonstrates consistent synchronization and mode-locking performance, essential for multi-MHz operation. (d) Optical spectra of the seed laser and post-external stage outputs of SPML-SS, with the output bandwidth measured at 102 nm at the 5-dB threshold. (e) Normalized unwrapped phase of fringe at a 4-mm optical path difference. The inset shows the residual curve from linear fitting, demonstrating an R-squared value of ~$1–10^{-6}$. This ensures accurate depth-resolved measurements in OCT imaging without the need for additional wavenumber linearization. (f) Temperature stress test at 15, 20, and 25℃ over 10 min. The optical power (71.3 mW ± 0.2 mW) showed < 0.3% variation, confirming thermal robustness and reliability of the laser system for clinical OCT imaging. OC: optical fiber coupler, SOA: semiconductor optical amplifier, EOM: electro-optic intensity modulator, CIR: optical fiber circulator, ISO: optical fiber isolator.

A dual CFBG architecture was integrated into the seed laser cavity. Comprising a long CFBG (lCFBG, Proximion, Sweden) and a short CFBG (sCFBG, TeraXion Inc., Canada), this configuration induced ultra-high dispersion variations and achieved near-zero net group delay during each round trip. The lCFBG, at an ultra-long length of ~10 m, was designed with a total group delay of ~93 ns, an average dispersion of approximately ±930 ps/nm, and a wavenumber-linear dispersion profile spanning a wavelength band (at the 3-dB threshold) from 1020 to 1120 nm. The sCFBG, operating across a bandwidth of 105 nm and centered at a wavelength of 1070 nm, minimized the chromatic dispersion of round trips in the fiber cavity across the operating bandwidth. The dispersion parameters D2, D3, and D4 of the sCFBG were precisely engineered for the 40.6-m PM980 fiber, with D2 at 1.594 ps/nm, D3 at -0.00389 ps/nm², and D4 at 0.00000435 ps/nm³. As shown in Fig. 2b, by implementing the sCFBG, the relative group delay of the net cavity was measured using the modulation phase-shift method [21], which was recorded at 3.8 ps (max-min) and exhibited a standard deviation of 1.3 ps across a bandwidth of 100 nm. This value guarantees stable mode locking across the entire laser bandwidth, even with modulation signals of pulse widths as short as a few tens of picoseconds.

The external stage allows additional wavenumber-linear time-stretching and efficient amplification, which doubles the data acquisition capacity and ensures sufficient optical power for high-sensitive imaging. Figs. 2c and 2d show the time-domain traces and optical spectra of the seed laser output and the post-external stage output of the SPML-SS, measured at Points A and B, respectively, as indicated in Fig. 2a. The output of the seed laser was coupled through the 10% output port of the optical coupler, operating with a pulse width of ~93 ns and a sweep rate of 5.0136 MHz, corresponding to a round-trip time of 199.457 ns and yielding a duty cycle of ~47%. This output passes through the external stage, gets reflected by the external lCFBG, and subsequently passes through the external SOA. The pulse width of the laser output was increased to 186 ns, corresponding to the total group delay of the two lCFBGs. These operations doubled the duty cycle to ~94%, promoted spectrum flattening with an output bandwidth of 102 nm (at the 5-dB threshold) owing to the gain saturation effect of the SOA, and increased the average output power to 120 mW. An optical fiber isolator (ISO, BPMCIR-1070-12-L-10-FA-F, OF-Link Communications Co., Ltd, China) was employed to block unwanted back reflections from the OCT system, ensuring

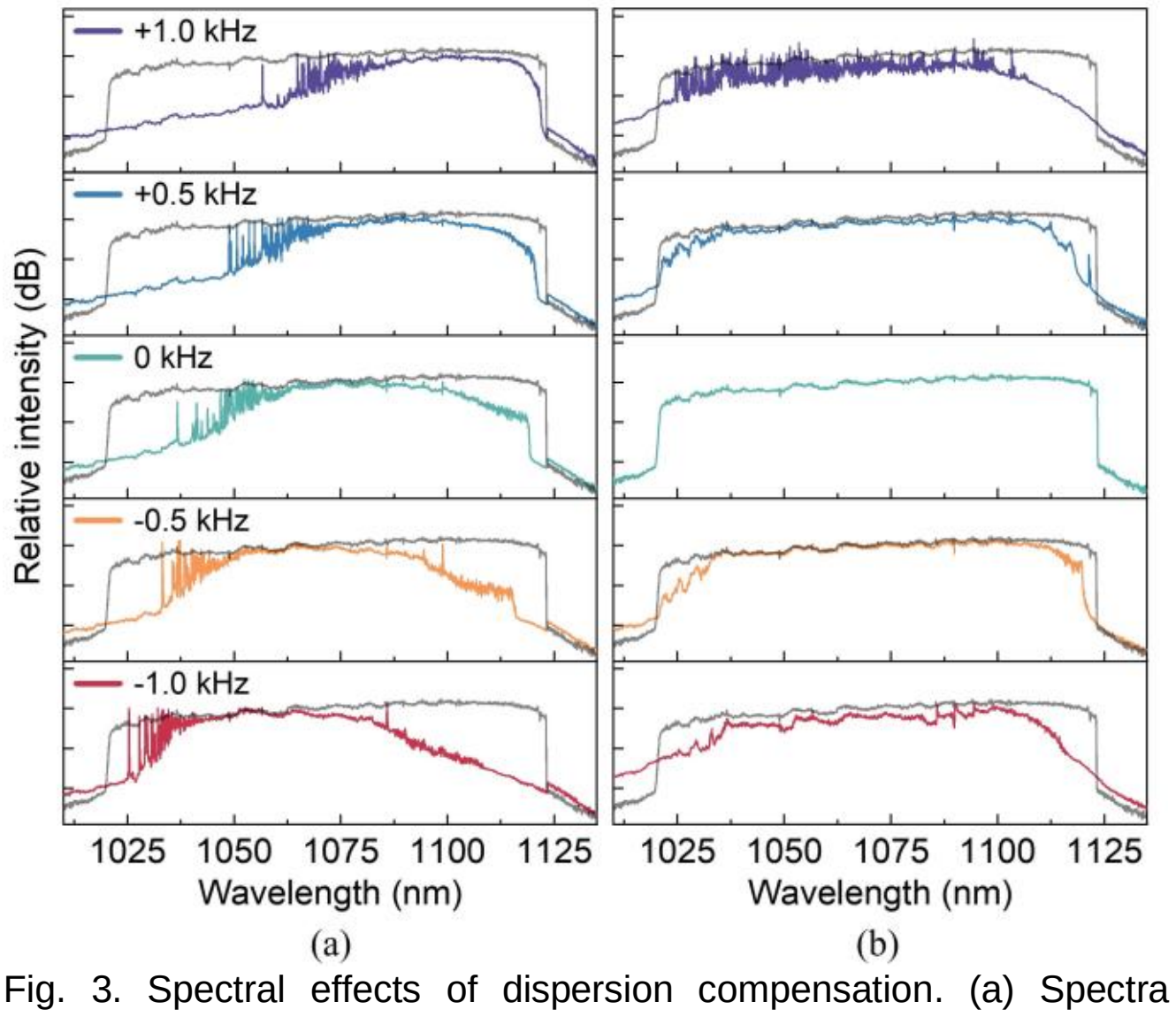

Fig. 3. Spectral effects of dispersion compensation. (a) Spectra recorded without dispersion compensation, showing significant variability and shifts toward longer wavelengths with increasing modulation frequency. These changes are due to the strong dispersion characteristic of the PM980 optical fiber. (b) Spectra recorded with dispersion compensation using an sCFBG. At zero offset modulation frequency, the spectrum shows a flat and broad profile across a 100-nm bandwidth, which serves as references for comparing other spectra represented by the gray baseline spectrum. Both (a) and (b) were measured at various modulation frequencies, adjusted in 500 Hz increments near the fundamental resonance frequency. Y-axis division = 5 dB.

stable operation.

The flat and broad spectrum ensures a high axial resolution and depth for imaging applications. As shown in Fig. 2e, the linear fitting analysis delivered an R-squared value of $1–10^{-6}$, indicating that the laser output maintained a highly linear-in-wavenumber characteristic over time. This is a crucial attribute for imaging applications, particularly OCT because it ensures the accuracy of depth-resolved measurements without requiring additional wavenumber linearization techniques [26].

As shown in Fig. 2f, the SPML laser underwent a temperature stress test at 15, 20, and 25 ℃ over a 10-minute interval. The average optical power was measured as 71.3 mW with a stability of ± 0.2 mW, corresponding to a variation of less than 0.3%. These measurements were conducted at Point B, located after the isolator, as indicated in Fig. 2a. Moreover, the output spectrum remained highly stable, with less than 0.5 dB changes across the tested temperatures. These results confirm the thermal robustness and consistent performance of the laser system across typical indoor temperatures, thereby ensuring the reliability of long-term OCT imaging in clinical settings.

Fig. 3 shows the effect of dispersion compensation on the laser output spectra, measured in 500 Hz increments of the modulation frequency near the fundamental resonance frequency using 68-ps electrical pulses. Stable mode locking occurs when the modulation frequency matches the fundamental resonance frequency of the laser cavity. Without dispersion compensation, as shown in Fig. 3a, significant spectral variations and a shift toward longer wavelengths occur because of the large negative dispersion of the PM980 optical fiber. In contrast, Fig. 3b presents the spectra with dispersion compensation using an sCFBG, resulting in a flat and wide profile across a 100-nm bandwidth at zero offset frequency. This spectrum, displayed as the gray baseline, serves as references for comparison. The y-axis division was 5 dB. These results indicate effective compensation for relative group delays within the laser cavity.

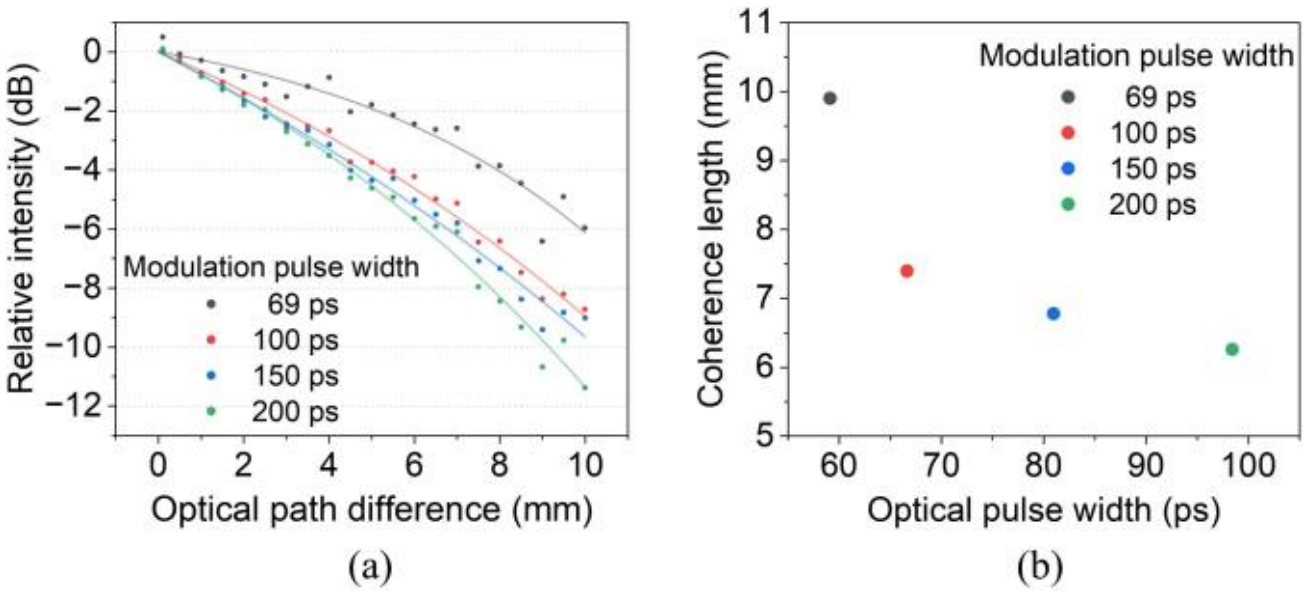

Fig. 4. Coherence characterization of dual CFBG SPML-SS. (a) Depth-dependent PSF roll-off curves measured for different modulation pulse widths (69, 100, 150, and 200 ps). (b) Coherence length as a function of optical pulse width, revealing an inverse relationship. Pulse widths as short as 69 ps correspond to a coherence length of ~9.9 mm, demonstrating the impact of pulse duration on coherence.

Fig. 4 illustrates the coherence characterization of the dual CFBG SPML-SS. In Fig. 4a, the depth-dependent point spread function (PSF) roll-off curves were measured using a Michelson interferometer with adjustable delays, where each curve corresponds to modulation pulse widths of 69, 100, 150, and 200 ps. Fringe signals were captured using a high-speed photoreceiver (ET-3500F, Coherent, Inc., USA; > 15 GHz bandwidth) and a rapid-sampling oscilloscope (DSA81304B, Agilent Technologies, USA; 40 GS/s sampling rate). The coherence lengths were derived from the -6 dB point of the PSF roll-off curves, demonstrating an inverse relationship with the optical pulse width. Fig. 4b shows this dependence by plotting the coherence length against the modulation pulse width. At the output of the electro-optical intensity modulator, a 69-ps electrical pulse generated a 59.2-ps optical pulse corresponding to a coherence length of 9.9 mm.

### C. Multi-MHz Phase-Sensitive Retinal Optical Coherence Tomography System

#### 1) OCT Interferometer

Fig. 5a shows a schematic representation of the in vivo human retinal OCT imaging system. A fiber-based Mach–Zehnder interferometer was designed to demonstrate the capabilities of the proposed SPML-SS for multi-MHz A-line rate and phase-sensitive retinal OCT. Four wideband optical fiber couplers (OCs, Thorlabs, USA) were incorporated, notably OC4, with a 10% tap in the sample arm to enhance the sensitivity of the OCT system in the 1.0-μm wavelength band. This design outperformed OCT systems that use traditional optical circulators in the same wavelength range, offering wideband operation, low-loss transmission (~0.5 dB), and negligible polarization-mode delay during signal detection. However, it requires high optical power from the SS owing to significant optical losses, which can reach up to 90%. Depending on the coupling ratio, this results in a tradeoff between the required optical power and system sensitivity. Despite these losses, the post-isolator output power is sufficient, delivering ~1.9 mW to the cornea, which is well below the ANSI standard for maximum permissible exposure [27]. In addition, this design allows balanced detection, which provides a strong suppression (~20 dB) of common-mode noise from the SPML-SS. To optimize the common-mode rejection in the 2.5-GHz balanced photodetector (BPD, PDB482C-AC, Thorlabs, USA), The optical path lengths of the OC3 output fibers were precisely balanced to below ~1 mm. The optical interference signals captured at the BPD were then digitized into sample data with 11-bit resolution using a 4 GS/s digitizer (ATS9373, Alazar Technologies Inc., Canada) incorporated into the computer. The axial PSF at multiple imaging depths was measured using a mirror target and linear stage, achieving an imaging depth of up to 2 mm and averaged axial resolution of 7.8 μm in the air (equivalent to 5.8 μm in water, given $n_{tissue}$ = 1.34). The sensitivity was measured to be 93.89 dB. To correct the refractive errors in the eye without movement of the OCT optics, an electrically focus tunable lens (A-58N1-P20, Corning Varioptic, USA) was incorporated after the collimator lens at the sample arm, which can adjust a diopter from -5 to 10. The 1/e$^2$ beam diameter at the pupil was 1.85 mm, corresponding to a 7.4-μm transverse resolution in the retina. To reduce the optical aberrations, two identical pairs of lenses, LP1 (100 mm × 2, AC254-100-C, Thorlabs, USA) and LP2 (75 mm × 2, AC254-075-C, Thorlabs, USA), were employed with effective focal lengths of ~50 and ~38 mm, respectively. A detailed 3D-rendered view of the optical imaging module is provided in Fig. 5b. It shows the components drawn to scale based on their actual size ratios in the system.

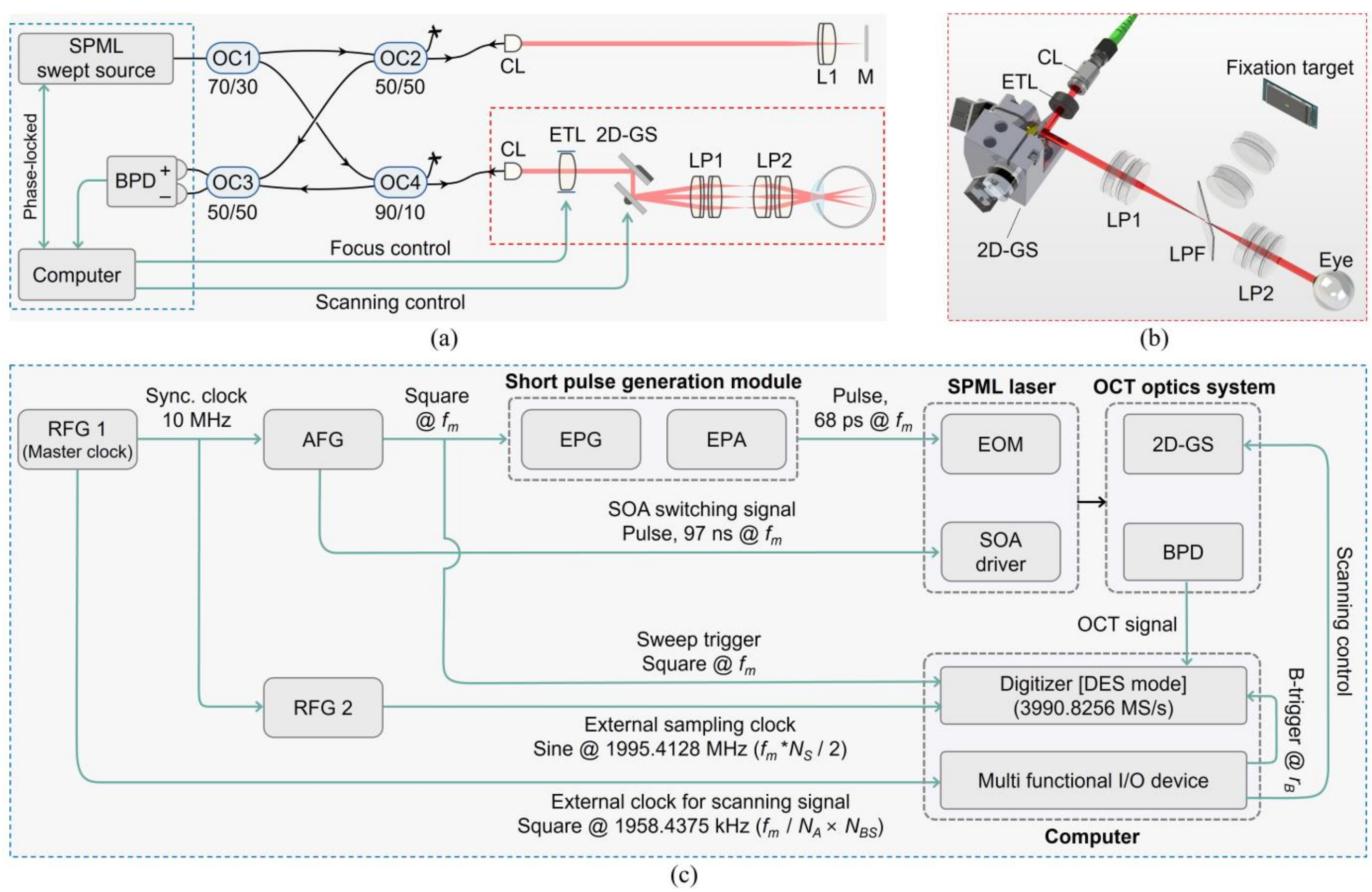

Fig. 5. Design and configuration of SPML-SS OCT system. (a) Schematic of the human retinal SS-OCT system. The red dashed box highlights the optical imaging module, detailed in (b), which is optimized for high-resolution and focus-tunable imaging of the human retina. The blue dashed box illustrates the phase-locking and signal acquisition system, expanded in (c), ensuring synchronized signal processing and phase stability. (b) 3D-rendered view of the optical imaging module, designed for high-speed OCT imaging with precise focus control and dynamic beam scanning. (c) Phase-locking and signal acquisition configuration. Phase-locking is achieved via radio-frequency signal generator (RFG) 1, which provides a master clock to synchronize the SPML-SS, scanning system, and digitizer, ensuring long-term and high precision phase stability essential for high-speed CV-OCTMA imaging. The arbitrary function generator (AFG) generates a modulation signal ($f_m$) of 5.0136MHz, matching the fundamental resonance frequency of the SPML laser cavity. This signal synchronizes the electrical pulse generator (EPG), which converts it into short pulses. These pulses are amplified by the electrical pulse amplifier (EPA) and delivered to the electro-optic intensity modulator (EOM). The digitizer, operating in dual-edge sampling (DES) mode at 3990.8256 MS/s, ensures fast and stable spectral sampling. SPML: stretched-pulse active mode-locked, OC: optical fiber coupler, CL: collimator, L: lens, M: mirror, BPD: balanced photodetector, ETL: electrically focus tunable lens, 2D-GS: dual-axis galvanometer optical scanner, LP: lens pair, LPF: longpass filter, SOA: semiconductor optical amplifier.

### 2) Optical Scanning Protocol

A dual-axis galvanometer optical scanner (2D-GS, Saturn 1B, ScannerMAX, USA) was used to perform two-dimensional (2-D) beam scanning of the retina. A bidirectional BM-scan protocol was used to achieve high-speed optical scanning while minimizing the number of unused data sections [2, 28, 29]. The fast-axis scanner was operated under the control of a sinusoidal waveform with a frequency of 1.95844 kHz. The slow-axis scanner repeatedly shifted stepwise depending on the direction of the fast-axis scanning, ensuring a consistent and extended time interval during each B-scan of the same location. To obtain high-quality CV-OCTMA images, eight B-scans were repeatedly collected at the same location, alternating between the odd and even B-scan locations by moving the slow-axis scanner. This procedure achieved an effective B-scan rate ($r_B$) of 3.91688 kHz and an interval of ~0.5 ms for repeated B-scans at the same location. Owing to the phase locking between the scanning system and the laser, a single B-scan yielded exactly 1280 A-line data. However, to mitigate the significant distortion in the edge portion of the sinusoidal scanning function used for fast-axis scanning, only 1000 A-line data points were utilized. For one volume imaging, 8000 B-scans were obtained for 1000 B-scan locations, rendering an imaging field of 4.1 × 4.4 mm$^2$ (1000 A-lines per B-scan × 1000 B-scan locations), completed within 2 s.

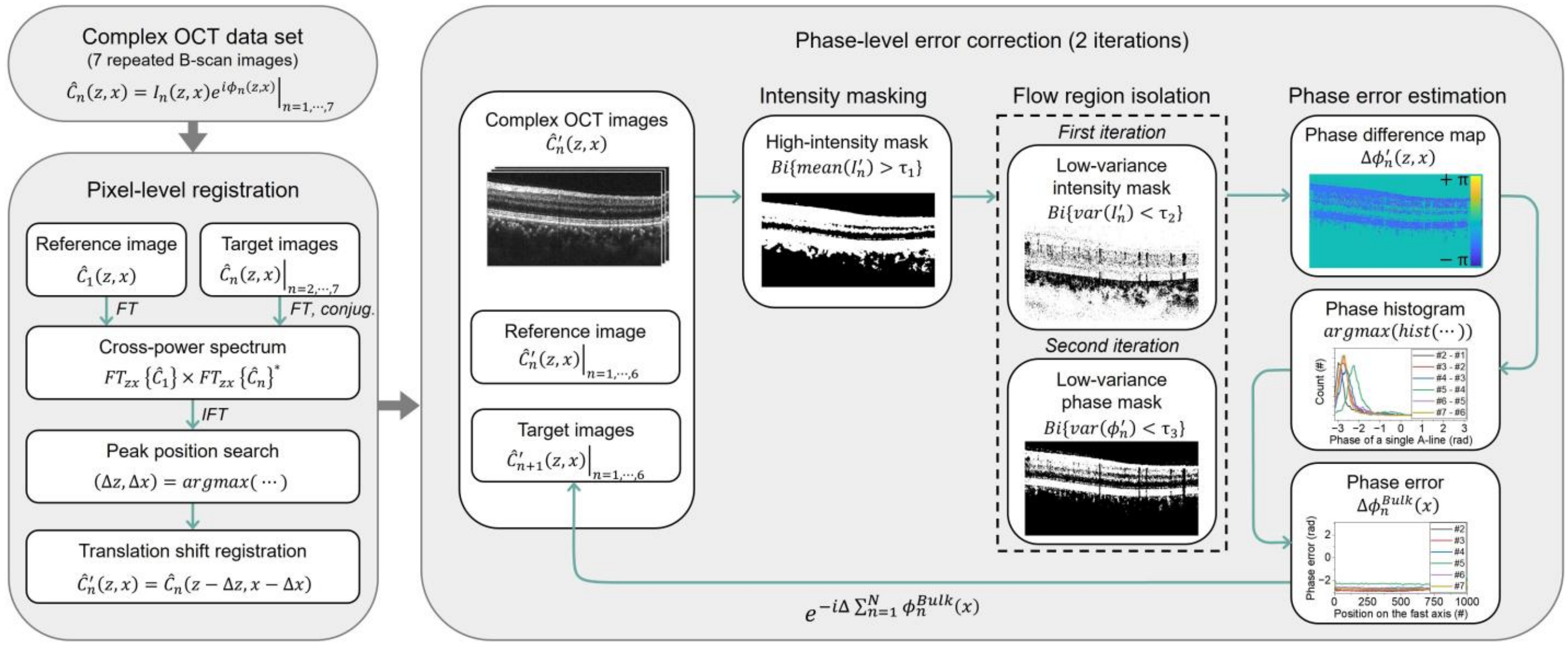

Fig. 6. Two-step bulk motion correction workflow. The workflow begins with pixel-level registration, aligning repeated B-scans through Fourier transform (FT)-based cross-power spectrum analysis and translation shift correction. Phase-level error correction follows, incorporating tissue masking based on a high-intensity mask to preserve anatomical structures, flow region isolation using low-variance intensity and phase masks across two iterations, and phase error estimation from static tissue regions via an averaged shifted histogram algorithm. This combined approach reduces motion artifacts, enhances alignment precision, and improves both the contrast and overall quality of 3D CV-OCTMA images. IFT: inverse Fourier transform.

### 3) Phase Locking Configuration and Signal Acquisition

Fig. 5c shows a phase-locking and signal acquisition configuration of the proposed multi-MHz retinal OCT system, illustrating the arrangement and interconnections of its various subsystems and components. Three signal generators were synchronized to a master clock in a radio-frequency signal generator (RFG) 1 (SG386, Stanford Research Systems, USA), which was based on a rubidium atomic time base to ensure high accuracy and temperature stability of phase locking. The system features a module for short electrical pulse signal generation, comprising an electrical pulse generator (EPG, EPG-220B-S-P-T-N-SF-N, Alnair Labs, Japan) and electrical pulse amplifier (EPA, DR-VE-10-MO, iXblue, France). The EPG produced an electric pulse train with a pulse repetition rate of 5.0136 MHz, aligned with the modulation frequency ($f_m$) from an arbitrary function generator (AFG, AFG31252, Tektronix, USA). A square waveform was used as the input signal to reduce additive jitter during short-pulse generation. After amplification by the EPA, the electrical pulse reached an amplitude of ~4.5 Vpp, which drove the EOM. The pulse width ranged from 69 ps to 225 ps, depending on the configuration. In synchronization with $f_m$, the switching signals of the SOA were generated by the AFG. These signals activated the SOA each time a stretched pulse passed through it. To select the maximum spectral bandwidth of the laser, the delay between the modulation signals of the EPG and SOA switching signals must be carefully set. Once the B-scan was initiated and the sweep trigger was activated, the high-speed digitizer continuously captured 1000 A-lines, with 796 samples per A-line ($N_s$), without requiring additional sweep trigger signals, thus allowing for a trigger jitter-free acquisition similar to that observed in spectral domain OCT. This operation was performed at a sampling rate of 3990.8256 GS/s using the dual-edge sampling mode. The process was externally synchronized by an RF signal, which operated at 1995.4128 MHz and originated from RFG 2 (68369A/NV, Anritsu, Japan). To synchronize the scanning control system, an external clock signal produced by RFG 1, was used, which was applied to a multifunctional I/O device (PCIe-6361, National Instruments, USA) on a computer. The control signal that operated the 2D-GS was externally synchronized with the signal from RFG 1, operating at a frequency of 1958.4375 kHz. This frequency was determined by dividing $f_m$ by the product of the number of samples per B-scan ($N_{BS}$) and A-line count per B-scan ($N_A$), which were set to 500 and 1280, respectively.

## D. Image Processing

A 12.5-GB data volume was captured within a field of view (FOV) of 4.1 × 4.4 mm$^2$, containing 8 million A-lines from 8000 B-scans. Each B-scan comprised 1000 fringe signals, with each signal containing 796 samples. Because of the sweeping duty ratio of 94%, only 748 samples from each A-line were utilized. After discarding one problematic B-scan, conventional

OCT preprocessing techniques, such as numerical dispersion compensation, spectral windowing, fast Fourier transform (FFT) with zero padding and fixed pattern noise removal [30], were applied, resulting in complex depth-encoded data with intensity and phase information up to a Nyquist depth limit of ~2 mm in air. The depth-encoded OCT signals of repeated B-scans at the same location can be expressed as

$$\hat{C}_n(z,x) = I_n(z,x)e^{i\phi_n(z,x)}, \tag{1}$$

where $\hat{C}_n(z,x)$ is the complex-valued OCT signal of a B-scan at specific $(z,x)$ positions, $I_n$ and $\phi_n$ are the intensity and phase values, respectively, and $n$ is the index of repeated B-scan frames in the same location.

Bulk motion artifacts from the imaging target or scanner affect the CV algorithms. As shown in Fig. 6, these artifacts were corrected sequentially, starting with pixel-level shift registration and progressing to the subpixel level using phase information [7, 31].

The repeated B-scans were aligned using a Fourier transform (FT)-based registration algorithm. This process involved transforming the images into the frequency domain via FFT, calculating the cross-power spectra, and applying an inverse Fourier transform (IFT). The peak in the transformed image indicates an optimal alignment shift.

$$\hat{C}'_n(z,x) = \hat{C}_n(z-\Delta z_n, x-\Delta x_n);\ \Delta z_1, \Delta x_1 = 0. \tag{2}$$

Here, $\Delta z_n$ and $\Delta x_n$ are the shift values of the n$^{th}$ frame of B-scan images in the same location along the z and x axes, respectively.

To enhance the contrast between static tissues and blood vessels, a two-step phase-level error correction workflow was employed, as illustrated in Fig. 6. This process began with tissue masking, which applied a high-intensity mask derived from a B-scan-averaged image to preserve the anatomical structures. Subsequently, a binary mask based on the intensity variance was used during the first iteration to suppress the scrambled phase information in large vessels and enhance the differentiation of static tissues and flow regions. In the second iteration, a binary mask based on the phase variance was applied to isolate the flow regions further and refine the phase corrections.

After rejecting the vessel region, the phase differences between adjacent B-scan images, denoted as $\Delta\phi(z,x)$, were computed according to the following equations and wrapped within a range of $\pm\pi$.

$$\phi'_n(z,x) = \angle\hat{C}'_n(z,x), \tag{3}$$

$$\Delta\phi'_{n+1}(z,x) = \phi'_{n+1}(z,x) - \phi'_n(z,x);\ \Delta\phi'_1(z,x) = 0. \tag{4}$$

An averaged shifted histogram algorithm was used for both iterations to estimate the bulk motion phase errors, specifically targeting the static tissue regions to enhance precision. The bulk motion phase errors, $\Delta\phi^{Bulk}(x)$, were estimated by identifying the phase value with the highest frequency in the histogram of phase differences for each A-line. The corrected complex OCT signal, $\hat{C}''_n$, was calculated as follows:

$$\hat{C}''_n(z,x) = \hat{C}'_n(z,x)e^{-i\sum_{n=1}^{N}\Delta\phi_n^{Bulk}(x)}. \tag{5}$$

After the two-step phase correction, CV-OCTMA images, $I_{CV}$, were generated using the following CV calculation as

$$I_{CV} = \frac{1}{N-1}\sum_{i=1}^{N}\left|\hat{C}''_i - \frac{1}{N-1}\sum_{j=1}^{N}\hat{C}''_j\right|^2. \tag{6}$$

Binary masks based on phase variance were applied to improve contrast and reduce noise. Moreover, the system-specific errors, such as image flipping and shifting, caused by the bi-directional BM-scan protocol, were compensated. The retinal data were segmented into four layers for targeted vascular analysis. Preprocessing techniques were applied to each section to reduce noise before generating the final 3D visualization, which was color-coded according to the depth of the 3D CV-OCTMA image. The total processing time for a single-volume data set on our prototype MATLAB (MathWorks, Inc., USA) platform was ~1 h and was performed on a desktop computer equipped with Windows 11 Pro x64, 64 GB RAM, and an Intel i9-12900K (3.2 GHz) CPU.

Fig. 7 highlights the effectiveness of the two-step phase-level correction workflow by showing the phase differences between adjacent B-scan images at three correction stages: original data, first correction, and second correction. The first correction method isolates static tissue regions using an intensity-variance-based mask to reduce motion artifacts. The second correction applies a phase-variance-based mask to further refine the phase estimation and improve the sensitivity to the residual flow regions. The results demonstrated a significant reduction in phase errors, with the mean squared error (MSE) decreasing from 9.9231 in the original data to 0.0018 after the first correction, and further to 0.0002 after the second correction. This workflow effectively minimized motion artifacts and ensured robust phase alignment across repeated B-scan datasets.

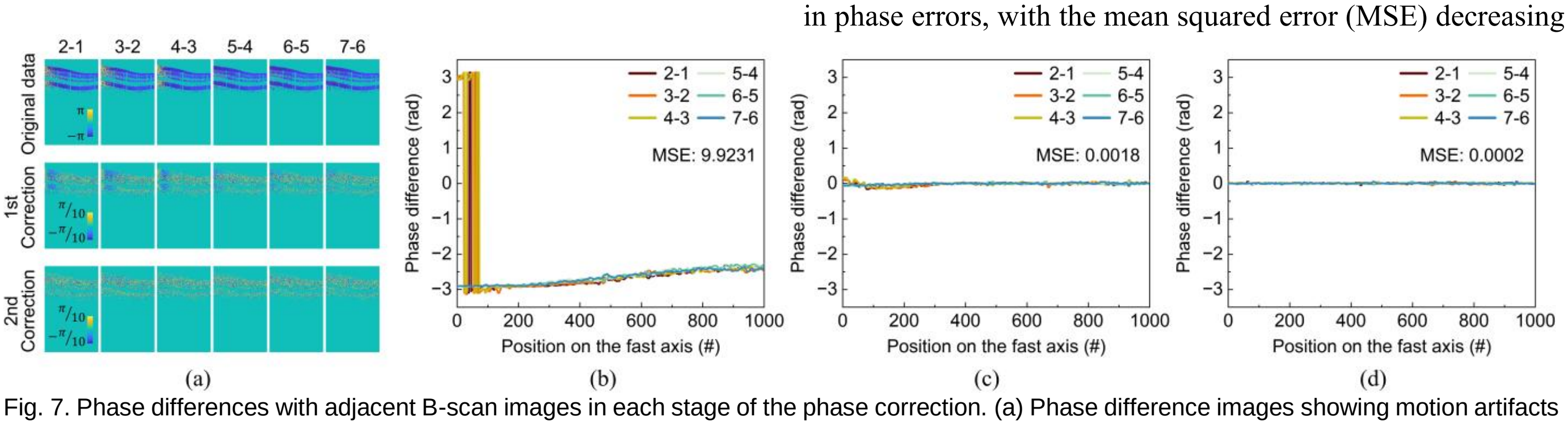

Fig. 7. Phase differences with adjacent B-scan images in each stage of the phase correction. (a) Phase difference images showing motion artifacts or instability in the optical beam scanner, calculated by comparing adjacent B-scan images (e.g., 2–1, 3–2, ..., 7–6). The top row represents the original phase difference data with a scale range from –π to π. The second and third rows depict the phase difference images after the first and second corrections, respectively, with a refined scale range of –π/10 to π/10. (b–d) Graphs of phase difference values along the fast axis for adjacent B-scan images. (b) Original phase difference data exhibits significant errors with a mean squared error (MSE) of 9.9231. (c) After the first correction, the MSE is reduced to 0.0018. (d) Following the second correction, the MSE further decreases to 0.0002.

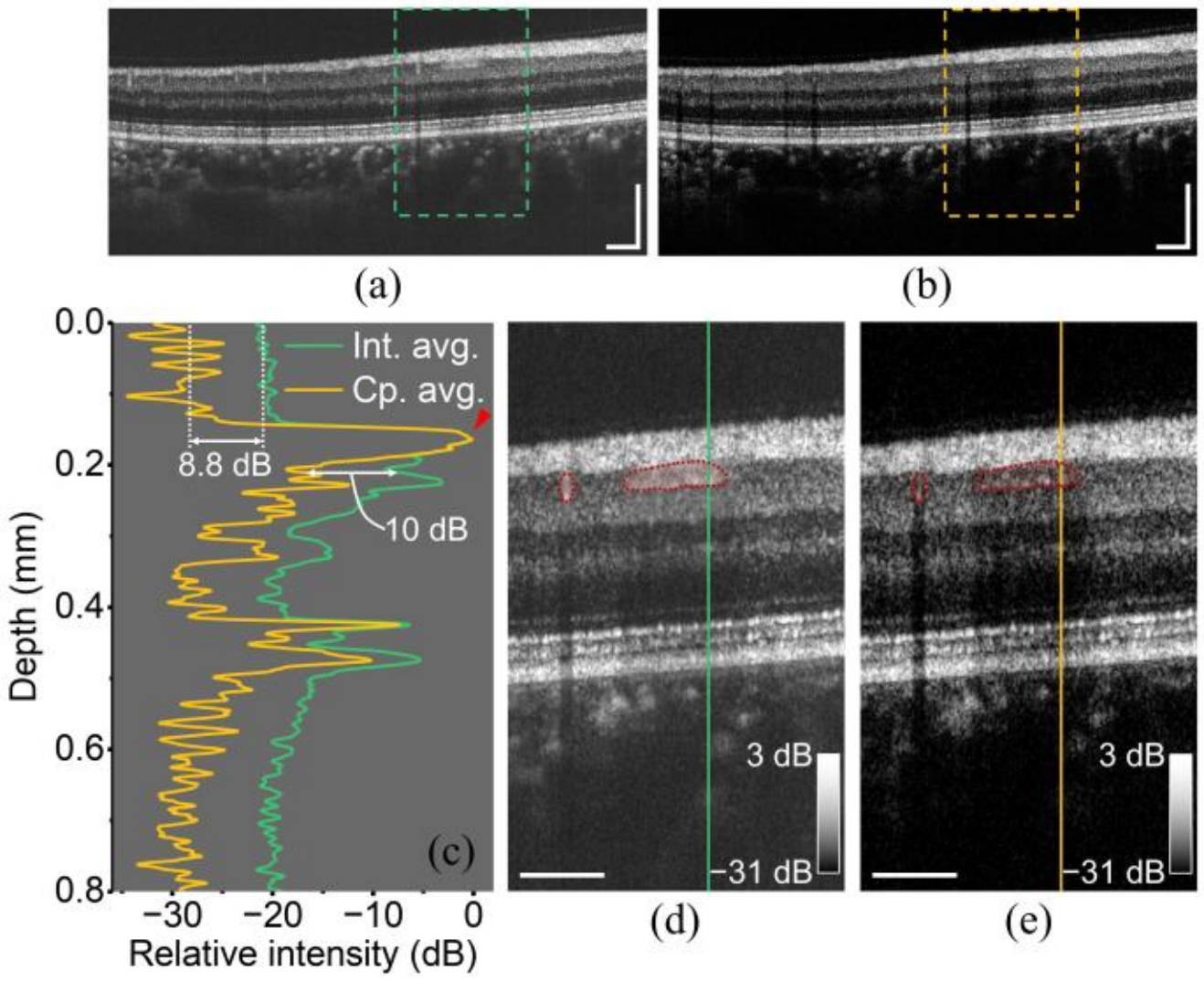

Fig. 8. Comparative averaging methods of *in vivo* human retina OCT image. (a, b) Cross-sectional images generated through two different averaging methods: (a) intensity-based and (b) complex-based averaging. Each image is derived from seven B-scan images taken at the same retinal location. (c) Relative intensity profiles of back-scattered signals along the depth (z-axis) for intensity- and complex-based average images, as indicated by the green and yellow lines in (d, e). These profiles show no signal depletion in the nerve fiber layer (red arrowhead), confirming the coherency of complex values across repeated B-scan images. (d, e) Magnified views of the areas within the dashed boxes in (a, b), with major blood vessels encircled in red dashed circles. Scale bar = 250 μm.

### E. *In Vivo* Human Retinal Imaging Protocol

Two healthy male volunteers (aged 31 and 34 years) participated in this study under protocols approved by the Pusan National University Institutional Review Board (IRB protocol number: PNU IRB/2023_07_HR). Written informed consent was obtained from all participants before imaging. The procedures performed in this study comply with the ethical standards of the Declaration of Helsinki. A pair of volumetric OCT images were procured from each volunteer, targeting two disparate regions of the retina: macula and optic disc. A fixation target aids in the precise acquisition of images from specific retinal sites. To minimize the influence of external ambient light [1, 32], all retinal imaging was performed in a dim environment without pharmacological pupil dilation.

## III. Results

### A. Complex-Based Averaged Image

The proposed ultra-high-speed, phase-stable imaging system, coupled with a meticulous error correction strategy, ensures the preservation of complex value coherency across successive B-scan retinal images under *in vivo* conditions. Figs. 8a and 8b were produced using intensity- and complex-based averaging, respectively. Both images were derived from an identical set of seven B-scan images, thereby facilitating a direct comparison of these distinct techniques. An in-depth analysis of the relative intensity profiles of the backscattered signals along the z-axis in both the intensity- and complex-based averaged images was conducted, as shown in Fig. 8c. These profiles, indicated by the green and yellow lines in Figs. 8d and 8e, respectively, show no signal depletion in the nerve fiber layer, as highlighted by the red arrowhead, affirming the coherency of the complex values across repeated B-scan images. A critical observation from this analysis was the significant reduction in background noise in the complex-based averaged image, quantified as 8.8 dB lower compared to its intensity-based counterpart. Consequently, the SNR was markedly improved in the stationary tissue regions depicted in the complex-based averaged image. Furthermore, the blood vessels appeared with a 10-dB lower intensity in the complex-based averaged image, a feature attributable to phase washout caused by dynamic blood flow. These differences are more pronounced in the magnified views of the selected areas in Figs. 8d and 8e. The major blood vessels appear darker in the complex-based average image. Notably, in the choroidal region comprising relatively large vessels, the strong phase changes induced by blood flow resulted in lower contrast in the complex-based averaged image. In conclusion, this comparative study demonstrated that complex-based averaging in OCT imaging is superior, particularly in applications requiring high sensitivity to stationary tissues.

### B. 3D Microvasculature Mapping

An *in vivo* CV-OCTMA system for the human retina based on the decorrelation of complex values across multiple repeated B-scans was presented. This study focused on the central macular region of the human retina (Fig. 9). The fundus photograph in Fig. 9a shows the specific area where the 3D complex OCT data were acquired, as indicated by the white box. The one-volume image with a 4.1 × 4.4 mm$^2$ FOV was captured in ~2.0 s. To investigate the structural details of this region, a retinal thickness map was constructed, as illustrated in Fig. 9b. This map, obtained from the 3D OCT image, represents the distance from the internal limiting membrane (ILM) to Bruch's membrane, providing a comprehensive view of retinal thickness variations within the area. CV-OCTMA was used to visualize the vascular architecture within this region. Fig. 9c shows an en-face MIP of the *in vivo* CV-OCTMA image, providing a detailed 3D view of the microvascular network within the central macular region. Here, the innermost boundary of the foveal avascular zone (FAZ), which the SVC surrounds, is clearly depicted. To aid in interpreting the image, the distance from the Bruch's membrane was encoded using a jet color map. This color-coding approach enables clear depth-resolved differentiation of the vascular structures within the retina. Notably, the vessel overlapping points, marked by red arrowheads in Fig. 9c, are clearly demonstrated, further highlighting the utility of this imaging approach.

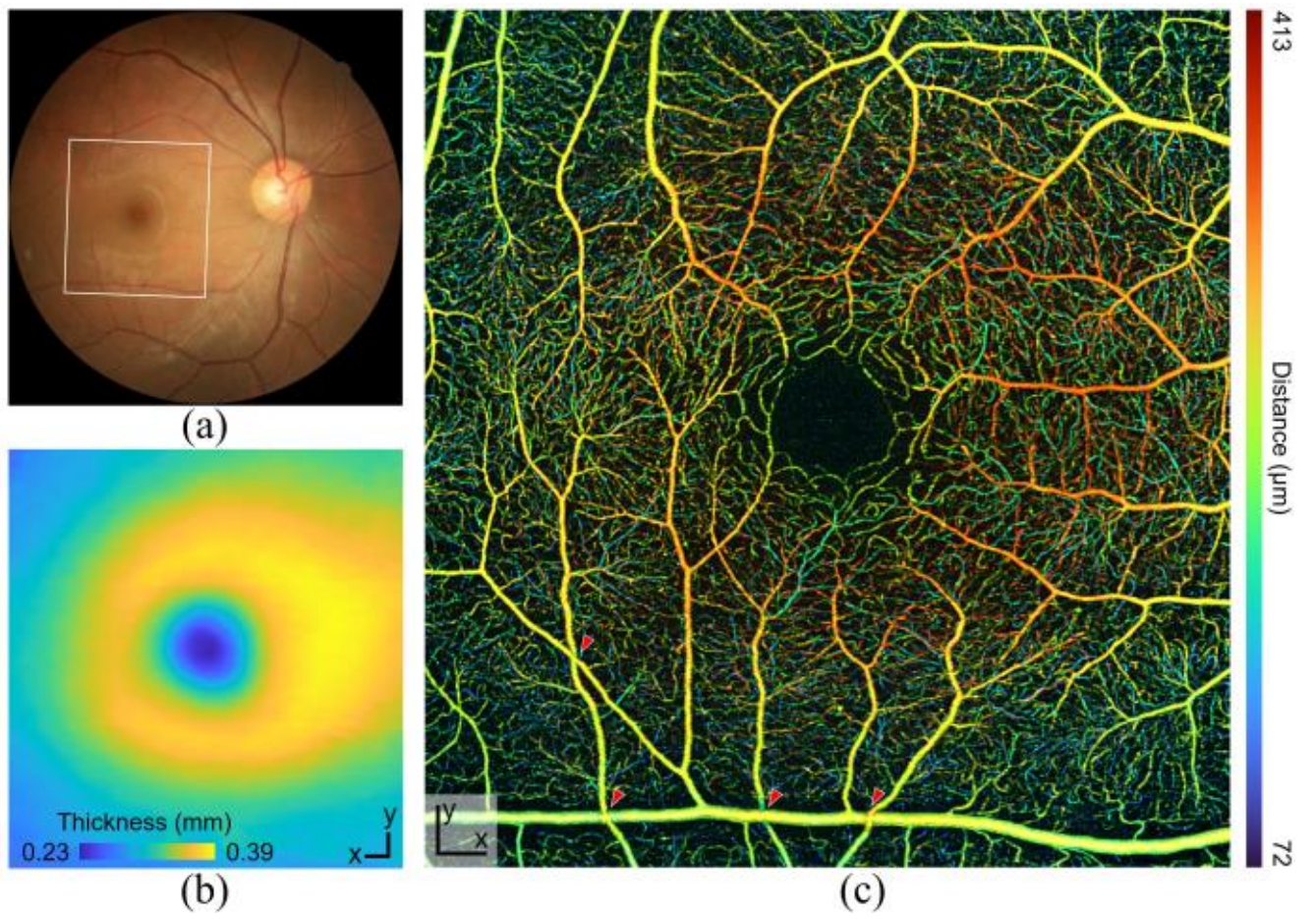

Fig. 9. Depth-encoded OCTMA image of the central macular region. (a) Fundus photograph, with a white box highlighting the region where the depth-encoded complex OCT data was acquired. (b) Retinal thickness map, visualizing the distance from the ILM to Bruch's membrane. (c) En-face MIP of the *in vivo* depth-encoded CV-OCTMA image within the central macular region of the human retina. This image vividly depicts the microvascular network, including the innermost boundary of the foveal avascular zone (FAZ), surrounded by the superficial vascular complex (SVC). Vessel overlapping points are marked by red arrowheads, clearly demonstrating the depth-resolved differentiation of vascular structures. The distance from the Bruch's membrane is coded using a jet color map. Scale bar = 250 μm.

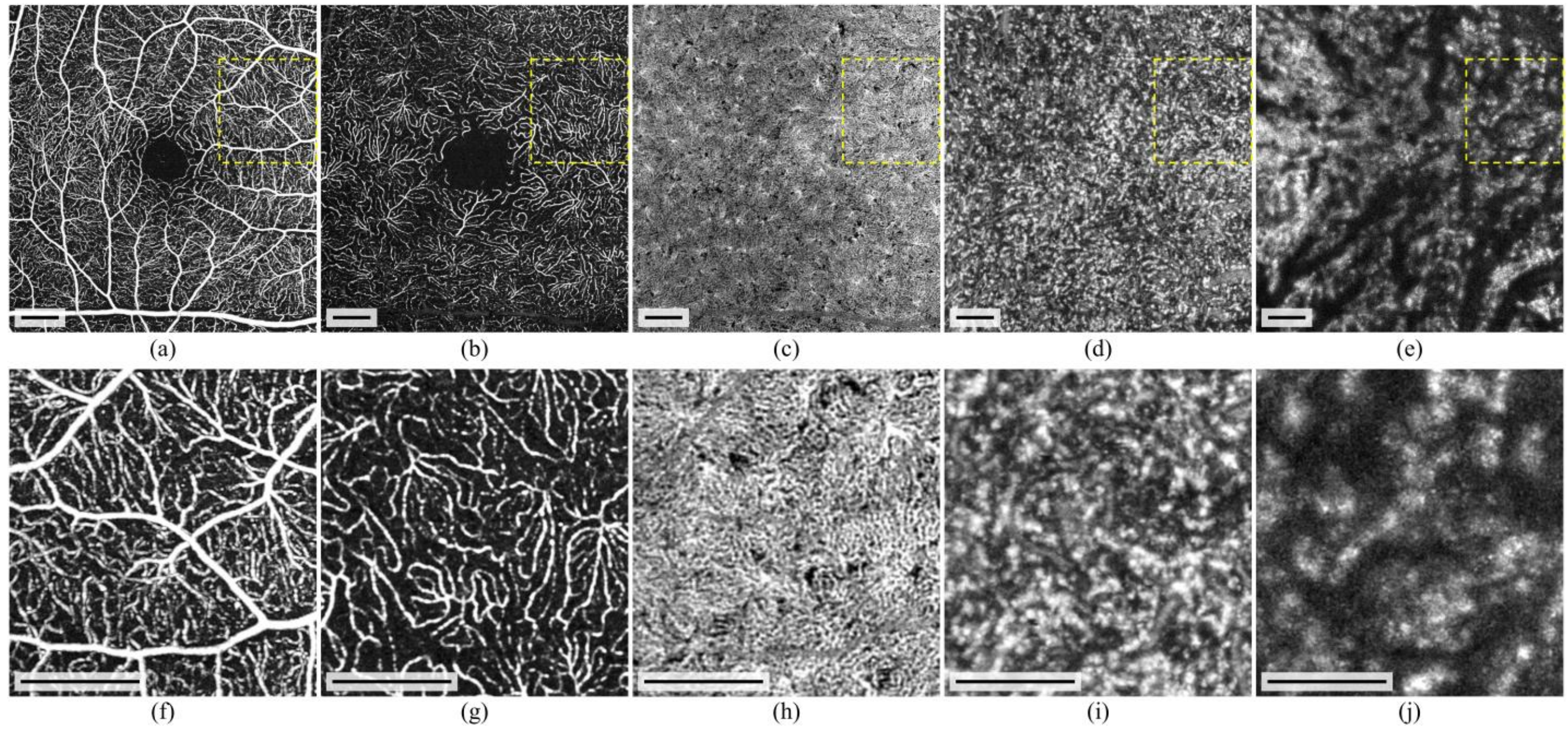

Fig. 10. High-resolution CV-OCTMA images of five different retinal layers. (a–e) En-face MIP images mapping the microvascular networks in each layer. (f–j) Zoomed-in views marked by yellow dashed squares in each corresponding upper image. (a, f) Superficial and intermediate capillary plexuses exhibit dense and complex microvascular networks. (b, g) Deep capillary plexus shows spider-like vortex patterns oriented toward the foveola, forming the innermost capillary networks. (c, h) Choriocapillaris (0–6 µm below Bruch's membrane) reveals its densely packed transverse microvasculature and thin vertical profile, effectively captured due to the high axial resolution of CV-OCTMA. (d, i) Sattler's layer (26–38 µm below Bruch's membrane) visualizes medium-sized choroidal vessels within the vascular lamina stroma. (e, j) Haller's layer (143–177 µm below Bruch's membrane) highlights large choroidal vessels appearing as dark stripe patterns. Scale bar: 500 µm.

### C. En-Face Projection Images of Five Different Retinal Layers

En-face MIP images were produced to capture high-resolution and wide-field microangiography images across different retinal layers (Figs. 10a–e). Each primary image, marked by a yellow dashed square, is complemented by a zoomed-in view (Figs. 10f–j), showing the detailed vascular networks within these layers.

Specifically, Fig. 10a reveals the dense and complex microvascular networks within the superficial and intermediate capillary plexuses (SCP and ICP). The images distinguish the dynamic blood flow from the surrounding static tissue, demonstrating the capability of CV-OCTMA to capture fine vascular details. Fig. 10b shows the microvasculature of the deep capillary plexus (DCP), with approximately eight spider-like vortex networks oriented toward the foveola, forming the innermost capillary networks.

The high average sampling density of ~2.5 ensured sub-Nyquist sampling, capturing even the smallest vessels across the entire FOV. To mitigate projection artifacts from the overlying larger vessels, a step-down exponential filter was applied to the 3D imaging data [33].

The choroidal layer images distinctly displayed the choriocapillaris, Sattler's, and Haller's layers (Figs. 10c–e). The choriocapillaris layer, characterized by its thin vertical profile and densely packed transverse microvasculature, is effectively depicted in Fig. 10h, located 0–6 µm below the Bruch's membrane. Owing to the high resolution of the proposed CV-OCTMA system (axially 5.8 µm, transversely 7.4 µm in tissue), an *in vivo* choriocapillaris image was captured, revealing intricate microvascular structures of the retina. Fig. 10d presents an en-face image derived from a 26–38 µm depth range below the Bruch's membrane. The stroma of the vascular lamina was prominently illuminated, and medium-sized choroidal vessels within the Sattler's layer were visualized. The large choroidal vessels characteristic of the Haller's layer (143–177 µm below the Bruch's membrane) are depicted as dark stripe patterns (Fig. 10e).

### D. Comparison of Angiography Algorithms for Retinal Microvasculature Visualization

To evaluate the performance of the angiography algorithms for visualizing the retinal microvasculature, three primary methods, —speckle variance (SV), phase variance (PV) and CV—were applied to the same 3D OCT dataset under identical post-processing conditions. A comparison was conducted across three retinal layers: the SCP + ICP, DCP, and choriocapillaris, as shown in Fig. 11.

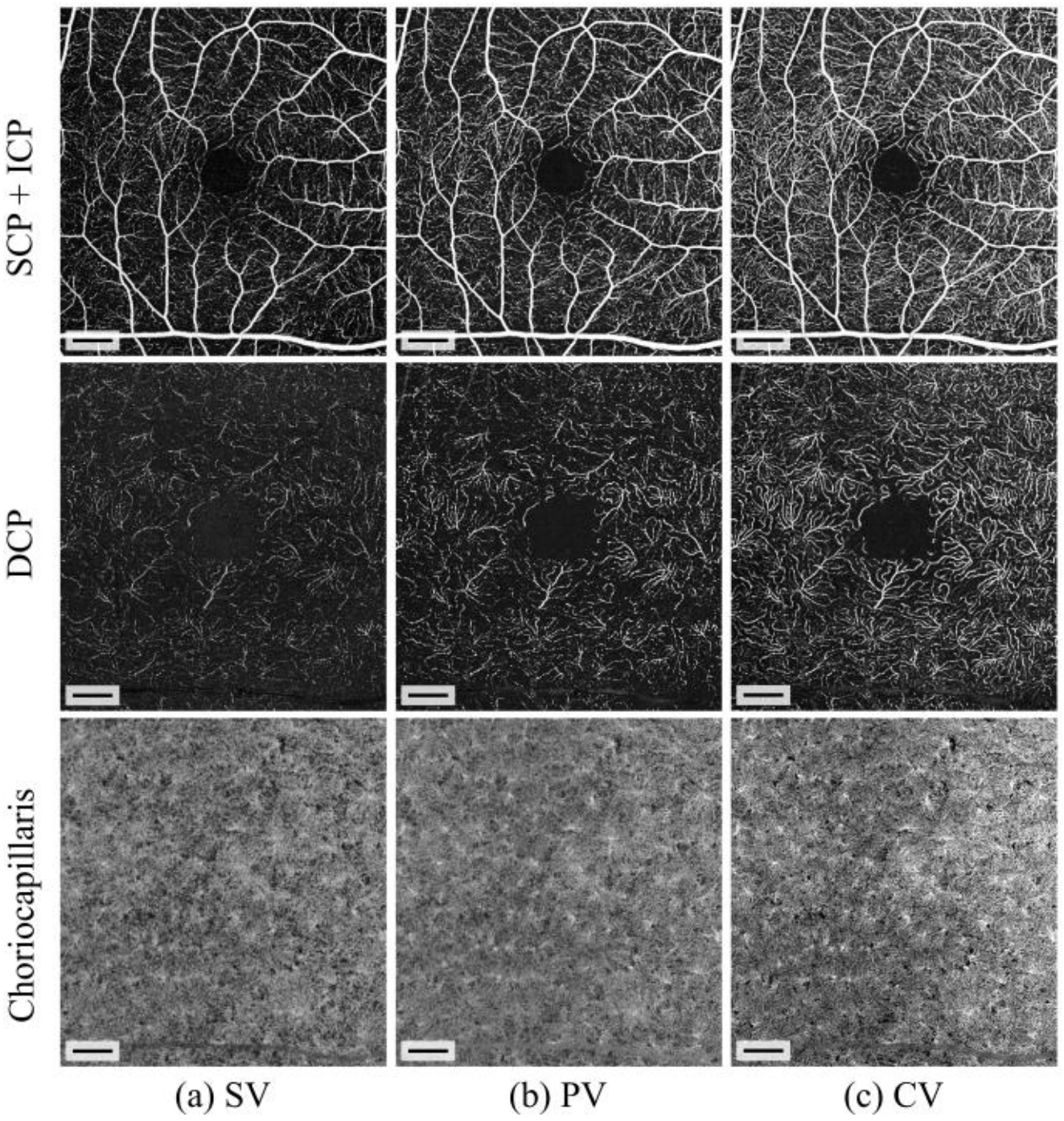

Fig. 11. Comparison of angiography algorithms for visualizing retinal microvasculature across three layers. The figure compares three angiography algorithms—(a) SV (speckle variance), (b) PV (phase variance), and (c) CV (complex variance)—using the same 3D OCT dataset and identical processing conditions. Rows correspond to different retinal layers: SCP + ICP (superficial and intermediate capillary plexuses), DCP (deep capillary plexus), and choriocapillaris. Each column illustrates how effectively the microvascular structures are visualized using the respective algorithm. The CV algorithm demonstrates superior visualization of fine vascular structures in all three layers, particularly in the DVC and choriocapillaris, where flow sensitivity and contrast are critical. Scale bars = 500 µm.

TABLE 1
SIGNAL-TO-NOISE RATIO (SNR) AND CONTRAST-TO-NOISE RATIO (CNR) COMPARISON OF ANGIOGRAPHY ALGORITHMS FOR VISUALIZING RETINAL MICROVASCULATURE

| | SCP + ICP | | DCP | |
|---|---|---|---|---|
| METHOD | SNR | CNR | SNR | CNR |
| SV | 10.9 | 6.7 | 8.1 | 2.3 |
| PV | 23.6 | 17.4 | 13.4 | 7.0 |
| CV | 24.4 | 20.6 | 13.9 | 9.0 |

The SV and PV algorithms are mathematically defined as

$$I_{SV} = \frac{1}{N-1}\sum_{i=1}^{N}\left|I_i'' - \frac{1}{N-1}\sum_{j=1}^{N} I_j''\right|^2, \quad (7)$$

$$I_{PV} = \frac{1}{N-1}\sum_{i=1}^{N}\left|\phi_i'' - \frac{1}{N-1}\sum_{j=1}^{N} \phi_j''\right|^2. \quad (8)$$

Here, $I_i''$ represents the intensity signal while $\phi_i''$ denotes the phase signal from the $i$-th repeated B-scan after applying bulk motion correction. $N$ corresponds to the total number of repeated B-scans used in the calculation.

To evaluate the quality of the angiography algorithms, the SNR and contrast-to-noise ratio (CNR) were calculated for each layer. The SNR and CNR are defined as

$$SNR = \frac{\langle S(x,y)\rangle_V}{\sigma_{BG}}, \quad (9)$$

$$CNR = \frac{\langle S(x,y)\rangle_V - \langle S(x,y)\rangle_{BG}}{\sigma_{BG}}, \quad (10)$$

where $\langle S(x,y)\rangle_V$ and $\langle S(x,y)\rangle_{BG}$ are the mean signal intensities in the vessel and background regions, respectively, and $\sigma_{BG}$ is the standard deviation of the background region.

The region of signals where the intensity masks were created was based on the microvasculature images across the SCP + ICP and DCP layers using the CV algorithm. The background region was selected at the center of macular images, corresponding to the FAZ, which is characterized by a dark region. The SNR and CNR values for the SV, PV, and CV algorithms in these layers are summarized in Table 1.

Table 1 highlights the superiority of the CV algorithm over SV and PV algorithms in the SCP + ICP and DCP layers. In the SCP + ICP layer, CV achieved the highest SNR (24.4) and CNR (20.6), significantly outperforming SV (SNR: 10.9, CNR: 6.7) and PV (SNR: 23.6, CNR: 17.4). Similarly, in the DCP layer, CV provided an improved SNR (13.9) and CNR (9.0) compared with SV and PV, indicating enhanced flow sensitivity and contrast in the deeper retinal layers. These quantitative results reinforce the qualitative findings presented in Fig. 11, where CV consistently provides clearer and more continuous vascular visualization across all layers.

### E. Quantitative and Qualitative Comparison of CV-OCTMA with Commercial OCTA Systems

The proposed CV-OCTMA system demonstrates significant advancements over two commercial OCTA systems—DRI Triton Angio (Topcon, Tokyo, Japan) and PLEX® Elite 9000 (Carl Zeiss Meditec, Dublin, CA, USA)—in both imaging quality and technical specifications. Fig. 12 highlights the qualitative differences in retinal microvasculature visualization across the three layers (SCP + ICP, DCP, and choriocapillaris). Table 2 summarizes the key performance metrics.

In standard OCTA procedures, volumetric scans require more than 5 s to complete. Scanning areas range from 2 × 2 mm² to 12 × 12 mm², however, larger fields of view degrade image quality owing to the fixed A-line rates of the OCT systems. To ensure a fair comparison of the image quality based on the maximum achievable lateral resolution of commercial devices, imaging was performed using a 3 × 3 mm² setting. The PLEX® Elite 9000, operating at 1,050 nm with an A-line rate of 100,000 A-lines/s, captures four successive B-scans for a 3 × 3 mm² image (300 A-lines, 300 B-scans) or two successive B-scans for a 6 × 6 mm² image (500 A-lines, 500 B-scans). In contrast, the DRI Triton maintained a consistent resolution of 320 A-lines and 320 B-scans across all areas at the same A-line rate, capturing four successive B-scans. The PLEX® Elite 9000 employs the optical microangiography (OMAG) algorithm, whereas the DRI Triton utilizes the OCT angiography ratio

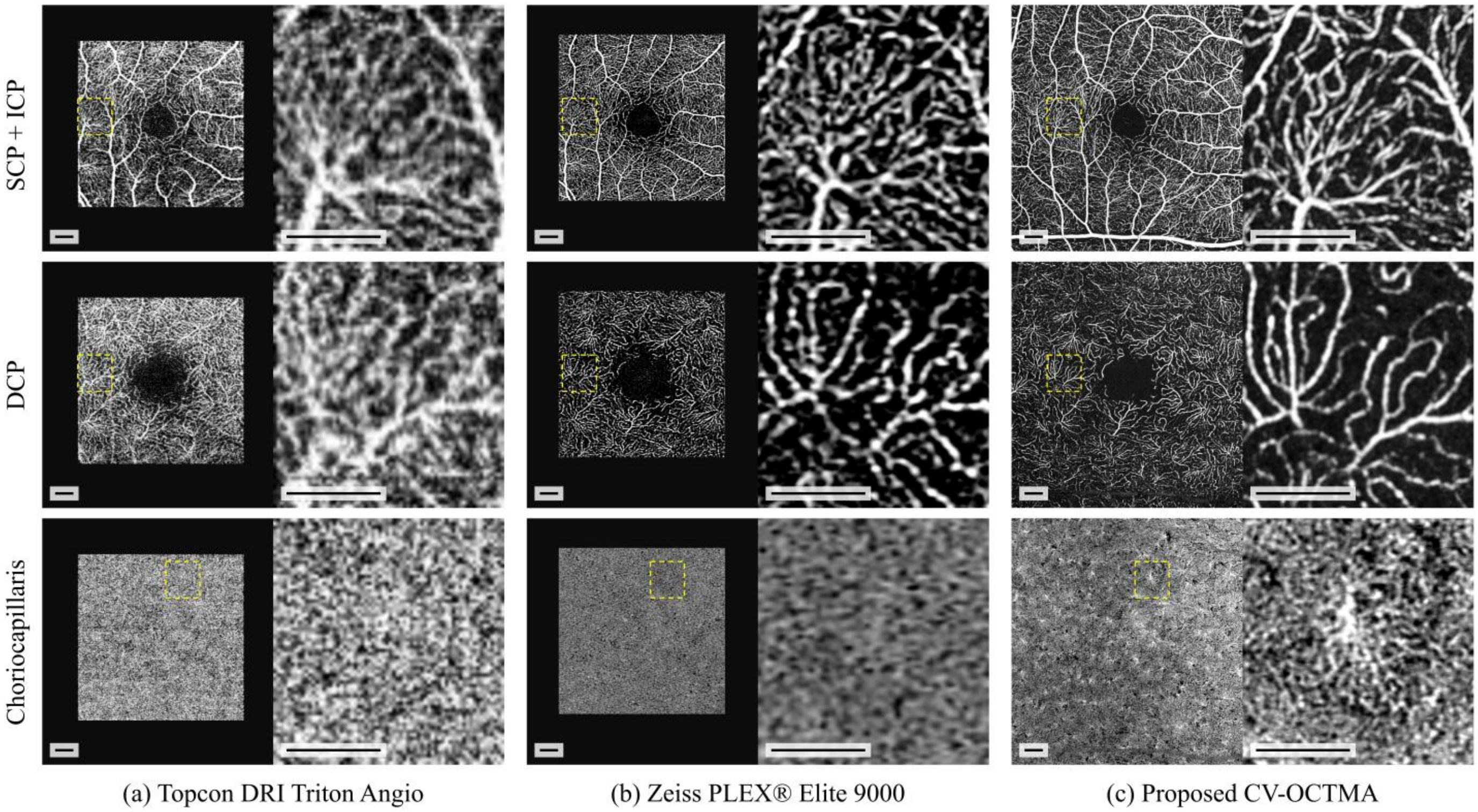

Fig. 12. Comparative retinal microvasculature imaging using three angiography systems. Rows correspond to SCP + ICP (superficial and intermediate capillary plexuses), DCP (deep capillary plexus), and choriocapillaris layers, while columns represent results from (a) Topcon DRI Triton Angio (3 × 3 mm$^2$ field of view), (b) Zeiss PLEX® Elite 9000 (3 × 3 mm$^2$ field of view), and (c) the proposed CV-OCTMA (4.1 × 4.4 mm$^2$ field of view). The CV-OCTMA demonstrates superior visualization of fine vascular structures, continuous capillary networks, and intercapillary spaces, particularly in the DCP and choriocapillaris layers. The yellow dashed squares indicate zoom-in areas for detailed visualization. Its enhanced flow sensitivity and reduced noise enable the precise identification of slow-flowing vasculature and anatomical details. Scale bar = 250 μm.

analysis (OCTARA) algorithm [34, 35].

In the SCP + ICP layer, CV-OCTMA (4.1 × 4.4 mm² field of view) provided seamless capillary visualization with reduced background noise, significantly outperforming DRI Triton Angio (3 × 3 mm² field of view) and PLEX® Elite 9000 (3 × 3 mm² field of view), which showed poor resolution and incomplete capillary networks. In the DCP layer, CV-OCTMA demonstrated superior spatial resolution and contrast, resolving fine vascular loops and vortex networks. In contrast, both commercial systems were unable to adequately define intercapillary spaces. In the choriocapillaris layer, CV-OCTMA achieved higher contrast and reduced noise, capturing intricate vascular patterns and flow dynamics, whereas DRI Triton Angio and PLEX® Elite 9000 produced ambiguous results.

As summarized in Table 2, CV-OCTMA delivered a larger field of view (4.1 × 4.4 mm²), with a pixel density of 243.9 × 227.3 pixels/mm, significantly higher than DRI Triton Angio (106.7 × 106.7 pixels/mm) and PLEX® Elite 9000 (100 × 100 pixels/mm). Despite its larger field of view, CV-OCTMA completed the scans in 2 s, outperforming the DRI Triton Angio (~5 s) and PLEX® Elite 9000 (~6 s). Furthermore, CV-OCTMA achieved superior axial (5.8 μm) and lateral (7.4 μm) resolution, surpassing DRI Triton Angio (axial: 8.0 μm, lateral: 20.0 μm) and PLEX® Elite 9000 (axial: 6.3 μm, lateral: 20.0 μm).

These results highlight the exceptional capabilities of CV-OCTMA in terms of imaging resolution, flow sensitivity, and anatomical detail. Compared with state-of-the-art OCTA systems, CV-OCTMA provides superior visualization of fine vascular structures, slow-flowing vasculature, and complex retinal details, emphasizing its clinical potential for diagnosing and monitoring retinal diseases.

## IV. Discussion

The proposed phase-locked time-stretch-based CV-OCTMA system achieved significant advancements in retinal imaging. It addressed the limitations of conventional OCTA methods through high-speed imaging, improved phase stability, and enhanced spatial resolution. With an A-line rate of 5 MHz and a phase stability of 4.8-mrad, the system enables precise visualization of intricate vascular structures, including the inner retina and choriocapillaris layers. In addition, the 3D mapping capability of CV-OCTMA overcomes the limitations of 2D en-face projection methods by distinguishing overlapping vascular structures. This unique advantage facilitates the development of 3D biomarkers, offering clinicians a more detailed and informative view of the retinal microvasculature [36]. These features are crucial for detecting and monitoring of retinal and systemic diseases, such as age-related macular degeneration, glaucoma, and diabetic retinopathy in the elderly [37], and Alzheimer's disease [38], where detailed vascular architecture analysis is essential.

Campbell *et al*. proposed a hammock structure for the retinal capillary plexus [39]. Unlike upper capillary networks, the DCP forms a collateral circulation with spider-like vortex capillaries that traverse the horizontal raphe [40]. This structure is an alternative pathway that redirects blocked flow in pathological conditions like retinal vein occlusion [41]. Several parafoveal vortex networks encompassing the foveola are discernible at the DCP level. These capillary networks are interconnected and constitute the FAZ of the DCP. As shown in Fig. 10b, the proposed CV-OCTMA precisely mapped the DCP of the human macular region, revealing enhanced brightness indicative of higher erythrocyte flux compared to the vortex capillary networks in the perifoveal region.

TABLE 2
COMPARISON OF MACULAR IMAGING SPECIFICATIONS BETWEEN COMMERCIALLY AVAILABLE SWEPT SOURCE SYSTEMS AND THE PROPOSED CV-OCTMA

| | TOPCON DRI TRITON ANGIO | | Zeiss PLEX® Elite 9000 | | Proposed CV-OCTMA |
|---|---|---|---|---|---|
| Field of view (mm × mm) | 3 × 3 | 6 × 6 | 3 × 3 | 6 × 6 | 4.1 × 4.4 |
| OCTA algorithm | OCTA-ratio analysis (OCTARA) | | Complex optical microangiography (OMAG) | | Complex-variance microangiography (CV-OCTMA) |
| Algorithm type | Amplitude | | Amplitude + Phase | | Amplitude + Phase |
| Light source (Center wavelength) | Swept source (1050 nm) | | Swept source (1040–1060 nm) | | Swept source (1070 nm) |
| A-line rate (Scanning speed) | 100,000 scans/s | | Normal mode: 100,000 scans/s<br>High-definition mode: 200,000 scans/s | | 5,013,600 scans/s |
| Interscan time | 24.0 ms | | 20.0 ms (4 B-scans) | 15.0 ms (2 B-scans) | 0.5 ms (7 B-scans) |
| Acquisition time | ~5 s | | ~6 s | ~7 s | 2 s |
| Axial imaging depth in tissue | 2.6 mm | | 3.0 mm | | 1.5 mm |
| Axial resolution in tissue | 8.0 μm | | 6.3 μm | | 5.8 μm |
| Lateral resolution | 20.0 μm | | 20.0 μm | | 7.4 μm |
| A-scan (scans) | 320 | 320 | 300 | 500 | 1,000 |
| B-scan (scans) | 320 | 320 | 300 | 500 | 1,000 |
| Horizontal size (mm) | 3.0 | 6.0 | 3.0 | 6.0 | 4.1 |
| Vertical size (mm) | 3.0 | 6.0 | 3.0 | 6.0 | 4.4 |
| Pixel density (H × V, pixels / mm) | 106.7 × 106.7 | 53.3 × 53.3 | 100.0 × 100.0 | 83.3 × 83.3 | 243.9 × 227.3 |
| Relative pixel density (H × V, %) | 43.7 × 46.9 | 21.9 × 23.5 | 41.0 × 44.0 | 34.2 × 36.7 | 100.0 × 100.0 |

Furthermore, detailed observation of the choriocapillaris layer in the central fovea using conventional OCTA systems is challenging because of limitations in lateral resolution and contrast [42]. However, the high resolution and efficient projection artifact rejection of CV-OCTMA enable *in vivo* visualization of the choriocapillaris region (Fig. 10h) at a level comparable to that of confocal microscopy images [43]. Notably, CV-OCTMA identifies feeding arterioles and interconnected capillary networks that form a spider web-like pattern, highlighting its potential as a non-invasive imaging modality for high-resolution visualization of the choriocapillaris.

Despite these advancements, the imaging depth of the current OCT configuration is restricted to 2 mm owing to the bandwidth limitations in photodetectors and digitizers. As signal detection technologies advance, this constraint is expected to be alleviated, enabling applications that require deeper imaging, such as intravascular or ultrawide-field retinal imaging. In addition, several challenges must be addressed prior to clinical implementation. This technology generates ~12.5 GB of raw OCT data per unit volume, substantially burdening on storage and processing. Reducing this to 3.6 GB after processing, using advanced compression algorithms such as JPEG 2000 and HEVC (H.265), improves efficiency. At the same time, implementing robust security protocols is critical to protecting patient data. Additionally, the current MATLAB-based prototype requires ~1 h to process a single volume, which is impractical for clinical workflows. Transitioning to graphics processing unit-based processing and automation can dramatically reduce the processing time and enhance operational efficiency. The current field of view, limited to 4 × 4 mm$^2$, is inadequate for imaging clinically significant regions such as the macula. Expanding this area to 6 × 6 mm$^2$ using resonant optical scanners would enable more effective diagnostics. Finally, the size and complexity of the system limit its portability and usability. A more compact design and an intuitive user interface are crucial for facilitating integration into clinical workflows and improving practicality.

Recent studies have investigated using visible-light sources in OCT for vascular imaging and $sO_2$ measurements, leveraging the sensitivity of hemoglobin absorption spectra [44, 45]. However, strong scattering and attenuation in the visible range and frequent spectral and motion artifacts in dynamic retinal environments limit their accuracy and repeatability [44, 45]. Photoacoustic imaging (PAI), a hybrid technique combining optical and ultrasound modalities, provides $sO_2$ mapping and structural information of deeper vasculatures but faces challenges in retinal imaging due to reduced SNR, limited axial resolution, and difficulty visualizing microvasculature, particularly *in vivo* [46, 47]. In contrast, the proposed CV-OCTMA system excels in high-resolution visualization of retinal microvascular networks, enabling rapid data acquisition and structural imaging. Integrating PAI with CV-OCTMA, similar to various previous integration attempts [46, 48], could combine their strengths—PAI for $sO_2$ mapping and deeper vasculature imaging, and CV-OCTMA for detailed microvascular visualization—offering a multidimensional vascular imaging approach that significantly advances diagnostics and research in clinical and experimental settings.

## Acknowledgment

The authors express their profound gratitude to the Late Professor Han Jo Kwon for his invaluable ophthalmological advice and extend their deepest condolences. The authors also acknowledge the technical support Huvitz Corporation (Republic of Korea) provided.

## References

[1] D. M. Sampson, A. M. Dubis, F. K. Chen, R. J. Zawadzki, and D. D. Sampson, "Towards standardizing retinal optical coherence tomography angiography: a review," *Light: Science & Applications,* vol. 11, no. 1, p. 63, 2022, doi: 10.1038/s41377-022-00740-9.

[2] J. Xu, S. Song, Y. Li, and R. K. Wang, "Complex-based OCT angiography algorithm recovers microvascular information better than amplitude- or phase-based algorithms in phase-stable systems," *Physics in Medicine and Biology,* vol. 63, no. 1, p. 015023, 2017, doi: 10.1088/1361-6560/aa94bc.

[3] A. S. Nam, I. Chico-Calero, and B. J. Vakoc, "Complex differential variance algorithm for optical coherence tomography angiography," *Biomedical Optics Express,* vol. 5, no. 11, p. 3822, 2014, doi: 10.1364/boe.5.003822.

[4] S. Yousefi, J. Qin, and R. K. Wang, "Super-resolution spectral estimation of optical micro-angiography for quantifying blood flow within microcirculatory tissue beds in vivo," *Biomedical Optics Express,* vol. 4, no. 7, pp. 1214-1228, 2013/07/01 2013, doi: 10.1364/BOE.4.001214.

[5] B. Považay *et al.*, "Impact of enhanced resolution, speed and penetration on three-dimensional retinal optical coherence tomography," *Optics Express,* vol. 17, no. 5, pp. 4134-4150, 2009/03/02 2009, doi: 10.1364/OE.17.004134.

[6] T. Klein and R. Huber, "High-speed OCT light sources and systems [Invited]," *Biomedical Optics Express,* vol. 8, no. 2, pp. 828-859, 2017/02/01 2017, doi: 10.1364/BOE.8.000828.

[7] J. V. Migacz, I. Gorczynska, M. Azimipour, R. Jonnal, R. J. Zawadzki, and J. S. Werner, "Megahertz-rate optical coherence tomography angiography improves the contrast of the choriocapillaris and choroid in human retinal imaging," *Biomedical Optics Express,* vol. 10, no. 1, pp. 50-65, 2019/01/01 2019, doi: 10.1364/BOE.10.000050.

[8] R. Poddar, J. V. Migacz, D. M. Schwartz, J. S. Werner, and I. Gorczynska, "Challenges and advantages in wide-field optical coherence tomography angiography imaging of the human retinal and choroidal vasculature at 1.7-MHz A-scan rate," *Journal of Biomedical Optics,* vol. 22, no. 10, pp. 1-14, Oct 2017, doi: 10.1117/1.JBO.22.10.106018.

[9] S. Chen *et al.*, "High speed, long range, deep penetration swept source OCT for structural and angiographic imaging of the anterior eye," *Scientific Reports,* vol. 12, no. 1, p. 992, 2022/01/19 2022, doi: 10.1038/s41598-022-04784-0.

[10] T. Klein, W. Wieser, L. Reznicek, A. Neubauer, A. Kampik, and R. Huber, "Multi-MHz retinal OCT," *Biomedical Optics Express,* vol. 4, no. 10, pp. 1890-908, 2013// 2013, doi: 10.1364/BOE.4.001890.

[11] Y. Miao *et al.*, "Phase-corrected buffer averaging for enhanced OCT angiography using FDML laser," *Optics Letters,* vol. 46, no. 16, p. 3833, 2021, doi: 10.1364/ol.430915.

[12] J. Zhang *et al.*, "Multi-MHz MEMS-VCSEL swept-source optical coherence tomography for endoscopic structural and angiographic imaging with miniaturized brushless motor probes," *Biomedical Optics Express,* vol. 12, no. 4, p. 2384, 2021, doi: 10.1364/boe.420394.

[13] Z. Wang *et al.*, "Cubic meter volume optical coherence tomography," *Optica,* vol. 3, no. 12, pp. 1496-1503, Dec 2016, doi: 10.1364/OPTICA.3.001496.

[14] I. Grulkowski *et al.*, "Retinal, anterior segment and full eye imaging using ultrahigh speed swept source OCT with vertical-cavity surface emitting lasers," *Biomedical Optics Express,* vol. 3, no. 11, pp. 2733-51, Nov 1 2012, doi: 10.1364/BOE.3.002733.

[15] W. Choi *et al.*, "Phase-sensitive swept-source optical coherence tomography imaging of the human retina with a vertical cavity surface-emitting laser light source," *Optics Letters,* vol. 38, no. 3, pp. 338-340, 2013/02/01 2013, doi: 10.1364/OL.38.000338.

[16] B. Johnson *et al.*, "Linewidth considerations for MEMS tunable VCSEL LiDAR," *Optics Express,* vol. 30, no. 10, p. 17230, 2022, doi: 10.1364/oe.456719.

[17] Y. Li, S. Moon, J. J. Chen, Z. Zhu, and Z. Chen, "Ultrahigh-sensitive optical coherence elastography," *Light: Science & Applications,* vol. 9, no. 1, 2020, doi: 10.1038/s41377-020-0297-9.

[18] B. J. Vakoc, S. H. Yun, J. F. De Boer, G. J. Tearney, and B. E. Bouma, "Phase-resolved optical frequency domain imaging," *Optics Express,* vol. 13, no. 14, p. 5483, 2005, doi: 10.1364/opex.13.005483.

[19] R. Khazaeinezhad, M. Siddiqui, and B. J. Vakoc, "16 MHz wavelength-swept and wavelength-stepped laser architectures based on stretched-pulse active mode locking with a single continuously chirped fiber Bragg grating," *Optics Letters,* vol. 42, no. 10, p. 2046, 2017, doi: 10.1364/ol.42.002046.

[20] B. Lee *et al.*, "Wide-field three-dimensional depth-invariant cellular-resolution imaging of the human retina," *Small,* vol. 19, no. 11, p. 2203357, 2023, doi: 10.1002/smll.202203357.

[21] T. S. Kim *et al.*, "9.4 MHz A-line rate optical coherence tomography at 1300 nm using a wavelength-swept laser based on stretched-pulse active mode-locking," *Scientific Reports,* vol. 10, no. 1, 2020, doi: 10.1038/s41598-020-66322-0.

[22] C. Wang *et al.*, "Polarization-isolated stretched-pulse mode-locked swept laser for 10.3-MHz A-line rate optical coherence tomography," *Optics Letters,* vol. 48, no. 15, pp. 4025-4028, 2023/08/01 2023, doi: 10.1364/OL.495786.

[23] S. Marschall, C. Pedersen, and P. E. Andersen, "Investigation of the impact of water absorption on retinal OCT imaging in the 1060 nm range," *Biomedical Optics Express,* vol. 3, no. 7, pp. 1620-1631, 2012/07/01 2012, doi: 10.1364/BOE.3.001620.

[24] S. Tozburun, C. Blatter, M. Siddiqui, E. F. J. Meijer, and B. J. Vakoc, "Phase-stable Doppler OCT at 19 MHz using a stretched-pulse mode-locked laser," *Biomedical Optics Express,* vol. 9, no. 3, p. 952, 2018, doi: 10.1364/boe.9.000952.

[25] A. Mahjoubfar, D. V. Churkin, S. Barland, N. Broderick, S. K. Turitsyn, and B. Jalali, "Time stretch and its applications," (in English), *Nature Photonics,* vol. 11, no. 6, pp. 341-351, Jun 2017, doi: 10.1038/nphoton.2017.76.

[26] H. Lee, G. H. Kim, M. Villiger, H. Jang, B. E. Bouma, and C.-S. Kim, "Linear-in-wavenumber actively-mode-locked wavelength-swept laser," *Optics Letters,* vol. 45, no. 19, pp. 5327-5330, 2020/10/01 2020, doi: 10.1364/OL.397715.

[27] A. N. S. Institute, *American National Standard for Ophthalmics - Light Hazard Protection for Ophthalmic Instruments* (American National Standard for Ophthalmics - Light Hazard Protection for Ophthalmic Instruments). New York: ANSI, 2016.

[28] T. Klein, W. Wieser, C. M. Eigenwillig, B. R. Biedermann, and R. Huber, "Megahertz OCT for ultrawide-field retinal imaging with a 1050 nm Fourier domain mode-locked laser," (in English), *Optics Express,* vol. 19, no. 4, pp. 3044-62, Feb 14 2011, doi: 10.1364/OE.19.003044.

[29] M. J. Ju, M. Heisler, A. Athwal, M. V. Sarunic, and Y. Jian, "Effective bidirectional scanning pattern for optical coherence tomography angiography," *Biomedical Optics Express,* vol. 9, no. 5, pp. 2336-2350, May 1 2018, doi: 10.1364/BOE.9.002336.

[30] S. Moon, S. W. Lee, and Z. Chen, "Reference spectrum extraction and fixed-pattern noise removal in optical coherence tomography," *Optics Express,* vol. 18, no. 24, pp. 24395-404, Nov 22 2010, doi: 10.1364/OE.18.024395.

[31] S. Makita, Y. Hong, M. Yamanari, T. Yatagai, and Y. Yasuno, "Optical coherence angiography," *Optics Express,* vol. 14, no. 17, pp. 7821-7840, 2006/08/21 2006, doi: 10.1364/OE.14.007821.

[32] G. Villatoro *et al.*, "Impact of Pupil Dilation on Optical Coherence Tomography Angiography Retinal Microvasculature in Healthy Eyes," *J Glaucoma,* vol. 29, no. 11, pp. 1025-1029, Nov 2020, doi: 10.1097/IJG.0000000000001647.

[33] B. J. Vakoc *et al.*, "Three-dimensional microscopy of the tumor microenvironment in vivo using optical frequency domain imaging," *Nature Medicine,* vol. 15, no. 10, pp. 1219-1223, 2009, doi: 10.1038/nm.1971.

[34] P. E. Stanga, E. Tsamis, A. Papayannis, F. Stringa, T. Cole, and A. Jalil, "Swept-source optical coherence tomography angio™ (Topcon Corp,

Japan): Technology review," (in eng), *Developments in Ophthalmology,* vol. 56, pp. 13-7, 2016, doi: 10.1159/000442771.

[35] R. K. Wang, "Optical microangiography: a label-free 3-D imaging technology to visualize and quantify blood circulations within tissue beds in vivo," *IEEE Journal of Selected Topics in Quantum Electronics,* vol. 16, no. 3, pp. 545-554, 2010, doi: 10.1109/JSTQE.2009.2033609.

[36] M. S. Sarabi *et al.*, "3D Retinal Vessel Density Mapping With OCT-Angiography," (in eng), *IEEE J Biomed Health Inform,* vol. 24, no. 12, pp. 3466-3479, Dec 2020, doi: 10.1109/jbhi.2020.3023308.

[37] E. Borrelli, D. Sarraf, K. B. Freund, and S. R. Sadda, "OCT angiography and evaluation of the choroid and choroidal vascular disorders," *Progress in Retinal and Eye Research,* vol. 67, pp. 30-55, 2018/11/01/ 2018, doi: https://doi.org/10.1016/j.preteyeres.2018.07.002.

[38] A. Katsimpris, A. Karamaounas, A. M. Sideri, J. Katsimpris, I. Georgalas, and P. Petrou, "Optical coherence tomography angiography in Alzheimer's disease: a systematic review and meta-analysis," *Eye,* vol. 36, no. 7, pp. 1419-1426, 2022, doi: 10.1038/s41433-021-01648-1.

[39] J. P. Campbell *et al.*, "Detailed vascular anatomy of the human retina by projection-resolved optical coherence tomography angiography," *Scientific Reports,* vol. 7, no. 1, p. 42201, 2017, doi: 10.1038/srep42201.

[40] P. L. Nesper and A. A. Fawzi, "Human parafoveal capillary vascular anatomy and connectivity revealed by optical coherence tomography angiography," *Investigative Ophthalmology & Visual Science,* vol. 59, no. 10, pp. 3858-3867, 2018, doi: 10.1167/iovs.18-24710.

[41] M. S. Kang, S. Y. Kim, S. W. Park, I. S. Byon, and H. J. Kwon, "Association between capillary congestion and macular edema recurrence in chronic branch retinal vein occlusion through quantitative analysis of OCT angiography," *Scientific Reports,* vol. 11, no. 1, 2021, doi: 10.1038/s41598-021-99429-z.

[42] A. Uji, S. Balasubramanian, J. Lei, E. Baghdasaryan, M. Al-Sheikh, and S. R. Sadda, "Choriocapillaris Imaging Using Multiple En Face Optical Coherence Tomography Angiography Image Averaging," *JAMA Ophthalmology,* vol. 135, no. 11, pp. 1197-1204, 2017, doi: 10.1001/jamaophthalmol.2017.3904.

[43] S. An *et al.*, "Effects of donor-specific microvascular anatomy on hemodynamic perfusion in human choriocapillaris," *Scientific Reports,* vol. 13, no. 1, p. 22666, 2023/12/19 2023, doi: 10.1038/s41598-023-48631-2.

[44] P. L. Nesper, B. T. Soetikno, H. F. Zhang, and A. A. Fawzi, "OCT angiography and visible-light OCT in diabetic retinopathy," *Vision Research,* vol. 139, pp. 191-203, 2017/10/01/ 2017, doi: https://doi.org/10.1016/j.visres.2017.05.006.

[45] I. Rubinoff *et al.*, "Adaptive spectroscopic visible-light optical coherence tomography for clinical retinal oximetry," *Communications Medicine,* vol. 3, no. 1, p. 57, 2023/04/24 2023, doi: 10.1038/s43856-023-00288-8.

[46] W. Liu and H. F. Zhang, "Photoacoustic imaging of the eye: A mini review," *Photoacoustics,* vol. 4, no. 3, pp. 112-123, 2016/09/01/ 2016, doi: https://doi.org/10.1016/j.pacs.2016.05.001.

[47] S. Jeon *et al.*, "In Vivo Photoacoustic Imaging of Anterior Ocular Vasculature: A Random Sample Consensus Approach," *Scientific Reports,* vol. 7, no. 1, p. 4318, 2017/06/28 2017, doi: 10.1038/s41598-017-04334-z.

[48] J. Park *et al.*, "Quadruple ultrasound, photoacoustic, optical coherence, and fluorescence fusion imaging with a transparent ultrasound transducer," *Proceedings of the National Academy of Sciences,* vol. 118, no. 11, p. e1920879118, 2021, doi: doi:10.1073/pnas.1920879118.